\documentclass[letterpaper,journal]{IEEEtran}
\usepackage{subcaption}
\usepackage{amsmath,amsfonts}
\usepackage{algorithmic}
\usepackage{algorithm}
\newtheorem{theorem}{Theorem}

\newtheorem{lemma}[theorem]{Lemma}
\usepackage{array}
\usepackage[caption=false,font=normalsize,labelfont=sf,textfont=sf]{subfig}
\usepackage{textcomp}
\usepackage{stfloats}
\usepackage{scalerel}
\usepackage{makecell}
\usepackage{url}
\usepackage{xcolor}
\usepackage{verbatim}
\usepackage{graphicx}
\usepackage[font=footnotesize]{caption}
\usepackage{mwe}
\usepackage{cite}
\DeclareMathOperator*{\argmin}{arg\,min}
\begin{document}

\title{RIS-Aided Spatial Nulling for Reflector Antennas: Algorithms, Analysis, and Nulling Limits}

\author{Xinrui Li,  R. Michael Buehrer,~\IEEEmembership{Fellow,~IEEE}, Steven W. Ellingson,~\IEEEmembership{Senior Member,~IEEE}
        % <-this % stops a space
\thanks{This work was supported in part by the National Science Foundation under Grant AST-2128506.}
\thanks{The authors are with the Wireless@VT, Bradley Department of ECE, Virginia Tech, Blacksburg, VA 24061 USA, E-mail: (lxinrui3@vt.edu; rbuehrer@vt.edu; ellingso@vt.edu).}% <-this % stops a space
}

\markboth{Journal of \LaTeX\ Class Files,~Vol.~14, No.~8, August~2021}%
{Shell \MakeLowercase{\textit{et al.}}: A Sample Article Using IEEEtran.cls for IEEE Journals}

\IEEEpubid{}

\maketitle

\begin{abstract}
Reconfigurable Intelligent Surfaces (RIS) have recently gained attention as a means to dynamically shape the wireless propagation environment through programmable reflection control. Among the numerous applications, an important emerging use case is employing RIS as an auxiliary mechanism for spatial interference nulling, particularly in large ground-based reflector antennas where sidelobe interference can significantly degrade the system performance. With the growing density of satellites and terrestrial emitters, algorithms with faster convergence speed and better performance are needed. This work investigates RIS-equipped reflector antennas as a representative example of RIS-assisted spatial nulling and develops algorithms for sidelobe cancellation at specific directions and frequencies under various constraints. For the continuous-phase case, we adapt the gradient projection (GP) and alternating projection (AP) algorithms for scalability and propose a closed-form near-optimal solution that achieves satisfactory nulling performance with significantly reduced complexity. For the discrete-phase case, we reformulate the problem using a penalty method and solve it via majorization–minimization, outperforming the heuristic methods from our earlier work. Further, we analyze the electric field characteristics across multiple interference directions and frequencies to quantify the nulling capability of the RIS-aided reflectors, and identify a simple criterion for the existence of unimodular weights enabling perfect nulls. Simulation results demonstrate the effectiveness of the proposed methods and confirm the theoretical nulling limits.
\end{abstract}


\section{Introduction}
The recent emergence of reconfigurable intelligent surfaces (RIS) has led to a wide range of applications for enhancing the performance of wireless communication systems. It typically consist of a planar array of sub-wavelength reflecting elements. By tuning the properties of these elements, RIS are able to modify the phase and amplitude of the incident signals, thereby controlling the propagation channels between the wireless transceivers \cite{8796365, 8741198}.\par
Motivated by the potential to control the wireless channel, extensive studies have explored the use of RIS for mitigating the interference across various communication scenarios ranging from traditional single-user and multi-user channels to large ground-based antenna systems. For example, single-user scenarios have been considered in \cite{8647620, 8683145, 8485924, 8815412}. Multi-user settings with inter-user interference have been addressed in \cite{8811733, 9001052}, where transmit beamforming at the base station and passive beamforming at the RIS are jointly optimized for maximizing the signal-to-interference-plus-noise ratio (SINR) or the achievable sum-rate. The authors in \cite{9405423} proposed a Riemannian manifold optimization approach to solve the problem of configuring the RIS coefficients to minimize the interference between all communicating pairs under constant modulus constraints. \par
Inspired by these advances, RIS can naturally serve as an auxiliary tool for spatial nulling in large ground-based antenna systems, which usually comprise large reflector antennas. For such instrument, the focused radiation pattern results in a strong main lobe gain that points toward the desired direction, but also produces sidelobes that could potentially receive the energy from unintended directions, exposing the system to potential interference. Traditional interference mitigation for reflector-based receivers often relies on time/frequency blanking at the receiver \cite{national2007handbook}, at the expense of data loss. Alternative solutions include the use of feed arrays \cite{monzingo2004introduction}, and reflector's edge treatments \cite{8013860} for facilitating sidelobe suppression. Unfortunately, these methods only shift the sidelobe peaks to other fixed positions and remain ineffective if the interference aligns with those peaks. \par
To overcome these limitations, the authors in \cite{10001662} proposed an approach that employs a RIS placed near a radio telescope to cancel interference before reception. The RIS, which receives the same incident wavefront as the telescope, cancels the interference at the receiver through creation of a destructive wavefront. Another approach which builds on both the concepts of edge treatments for suppressing the sidelobe levels and RIS has been proposed in \cite{9400735}. In this approach, the rim of a reflector antenna is equipped with reconfigurable elements which are capable of introducing a phase shift to the reflected electromagnetic waves. Owing to the ease of RIS deployment, an existing radio telescope can be retrofitted into a new one with minimal changes to the original reflector geometry and positioning hardware. Furthermore, the authors in \cite{10403767} presented a practical design for such a reconfigurable antenna and demonstrated the efficacy of this concept through full-wave simulation. \par
The rapidly increasing number of communication systems—particularly the explosive growth of LEO satellites for future satellite networks—highlights the increasing need for RIS-assisted spatial nulling as a flexible interference mitigation strategy. In this work, we build on the system proposed in \cite{9400735} as an representative example of the RIS-aided spatial interference suppression techniques and discuss new algorithms for determining the complex weights needed at each reconfigurable element to cancel sidelobes at specific angles and frequencies. 
Although this architecture is related to RIS-assisted interference mitigation, the resulting problem differs from conventional RIS beamforming. Existing RIS beamforming studies typically model the RIS as a planar auxiliary surface and optimize the RIS coefficients to improve communication metrics such as SINR, sum rate, or received signal power. In contrast, the system considered here is a partially reconfigurable reflector antenna: only a narrow annular region on the rim is controllable, while the majority of the reflector remains fixed. The weight-design problem is therefore dominated by the electric field contribution of the RIS elements for cancelling the interference response from the non-reconfigurable portion.  \par
In this paper, we consider three practically relevant weight-design regimes to evaluate the feasibility and achievable nulling performance of the proposed rim-loaded RIS reflector architecture. Within these regimes, we develop and adapt suitable weight-selection algorithms for different weight constraints, and compare the resulting performance with those presented in \cite{9886543, 10115669}. Specifically, we first derive the unconstrained optimal weights needed to cancel sidelobes at an arbitrary direction and frequency while maintaining the mainlobe gain in the least-squares sense, obtained via the pseudoinverse solution. 
In practice, RIS weights are often restricted to constant modulus to reduce hardware cost and implementation complexity. The resulting unit-modulus least-squares (UMLS) problem is non-convex and can be addressed by existing approaches such as semidefinite relaxation (SDR) \cite{5447068} or phase-only optimization \cite{771033, 1364142}. \par
However, these methods either increase the (already large) dimensionality of the problem or involve the calculation of a large matrix inverse, making them unsuitable for large-scale problems. Therefore, we adapt two projection-based methods—the gradient projection (GP) \cite{7849224, 10284537} and alternating projection (AP) \cite{escalante2011alternating} algorithms. They are promising solutions due to their low computational complexity and scalability. Moreover, we will show that for the problem of interest, although AP generally converges faster than GP, its final solution can be inferior in certain scenarios as compared to the solution provided by GP; thus, both algorithms offer distinct advantages in our problem. To further reduce complexity in finding the continuous-phase unimodular solution, we derive a closed-form expression for the approximate optimal unimodular weights by exploiting structural properties unique to our problem. This approximation achieves near-optimal performance while requiring significantly less computation time than any of the discussed iterative algorithms. \par
When the weights are further restricted to discrete-phase unimodular values, the problem becomes even more challenging as it reduces to finding an M-ary Phase Shift Keying (PSK) vector, which is generally hard to manipulate algebraically. In our previous work \cite{10115669}, this problem was addressed using serial search and simulated annealing, which is often used when the search space of an optimization problem is discrete. However, we observed that this approach suffers from degraded performance and slower convergence as the number of interference directions to null increases, and it also lacks reproducibility due to its inherently stochastic nature. To overcome these limitations, we adopt a penalty method to transform the problem into one with a convex constraint but non-convex objective function. This new problem can be tackled by the majorization minimization method \cite{10677490, 7547360}, which demonstrates superior performance compared to heuristic algorithms in terms of both computational time and nulling ability. \par
While much of the existing literature dealing with the UMLS problems pays significant attention to the minimization of the objective function, the conditions under which there exists a unimodular solution that minimizes the objective function to zero have not been fully investigated. The reason this matters is that for a large ground-based antenna, especially for the reflector antenna, the sidelobe gain towards the interference source can be highly sensitive to even small perturbations in the optimal weights, meaning that a slight increase in the objective function value may result in substantial degradation of the interference suppression. In this work, we examine the nulling limits of RIS-equipped reflector antennas which provides insight into the nulling capability of the RIS-aided antenna. This analysis is particularly relevant for antennas deployed in the dense spectral environments, where the presence of multiple interfering sources may challenge the achievable nulling performance of a RIS-assisted system. Specifically, by analyzing the correlation between the electric-field intensity vectors generated by the reconfigurable rim elements for different zenith angles, azimuth angles, and frequencies, we show that finding a unimodular solution becomes increasingly difficult as this correlation grows—either due to larger number or smaller angular separations of the interference sources. Moreover, by analyzing the properties of the unconstrained weights, we find a simple criterion for determining the existence of unimodular weights which strictly meet optimality criterion (minimizes the objective function to zero), and demonstrate the validity of this criterion through simulations. \par
An earlier conference version of this work appeared in \cite{10773626}. Relative to \cite{10773626}, this longer version also discusses the AP algorithm for obtaining continuous-phase unimodular solutions and provides a more detailed analysis of the electric field intensity correlation across multiple directions (and frequencies) for a reflector antenna system, which directly impacts the nulling performance of the algorithm. In addition, we present a more extensive set of simulation results to validate our theoretical analysis and to assess the performance of the proposed methods under a broader set of scenarios. To the best of our knowledge, the weight optimization of RIS elements mounted on the rim of a dish reflector antenna for adaptive sidelobe nulling and the corresponding nulling limits characterized by the correlation of the electric field intensity have not been systematically studied before. \par
The remainder of this paper is organized as follows: Section~\ref{sec:II} describes the system model used in this work. Section~\ref{sec:III} presents the methods for determining the optimal weights under various constraints. Section~\ref{sec:IV} explores the properties of the electric field intensity for different directions and frequencies, and establishes the limits of nulling performance for the reconfigurable dish. Section~\ref{sec:V} provides numerical results which characterize the performance of the discussed algorithms. Subsequently, Section~\ref{sec:VI} concludes the paper. 

\textit{Notation}: Throughout the article, we use the following notation: \(\hat{\mathbf{x}}\), \(\hat{\mathbf{y}}\), and \(\hat{\mathbf{z}}\) are the orthonormal basis of the Cartesian coordinate system, ``\(\cdot\)" and ``\(\times\)" denote the dot product and cross product between two position vectors, respectively. \(\mathbf{x}\) is a vector, \(\mathbf{X}\) is a matrix, \(\mathcal{X}\) is a set, \(\mathrm{conv}(\mathcal{X})\) is the convex hull of \(\mathcal{X}\), a vector of all ones is denoted \(\mathbf{1}\), \(\mathbb{R}\) and \(\mathbb{C}\) are the set of all real and complex numbers, respectively. \((\cdot)^T\) and \((\cdot)^H\) denote the transpose and Hermitian transpose, respectively. \(\mathrm{diag}(\mathbf{x})\) denotes the \(n \times n\) diagonal matrix which has diagonal entries \(x_1, \dots, x_n\). \(\lVert \cdot \rVert_p\) and \(\lVert \cdot \rVert_F\) denote the \(\ell_p\) norm and Frobenius norm, respectively, \(\lVert \cdot \rVert\) denotes the Euclidean norm. \(\Re(x)\) and \(\Im(x)\) denote the real and imaginary parts of \(x\), respectively. Given a matrix \(\mathbf{X}\), \(\lambda_i(\mathbf{X})\) denotes the \(i\)-th largest eigenvalue of \(\mathbf{X}\).
\begin{equation}
\Pi_{\mathcal{X}}(\mathbf{x}) = \argmin_{\mathbf{y} \in \mathcal{X}} \lVert \mathbf{x} - \mathbf{y}\rVert^2
\end{equation}
denotes the projection of \(x\) onto \(\mathcal{X}\). Function \(f\) is said to be Lipschitz continuous gradient on \(\mathcal{X}\) if there exists a constant \(L\) such that 
\begin{equation}
\lVert \nabla f(\mathbf{x}) - \nabla f(\mathbf{y}) \rVert \leq L \lVert \mathbf{x} - \mathbf{y} \rVert , \quad \forall \ \mathbf{x},\mathbf{y} \in \mathcal{X},
\end{equation}
and \(L\) is the Lipschitz constant of \(\nabla f\) on \(\mathcal{X}\).\par

\section{System Model}
\label{sec:II}
\begin{figure}[t!]
\begin{center}
\includegraphics[scale=0.5]{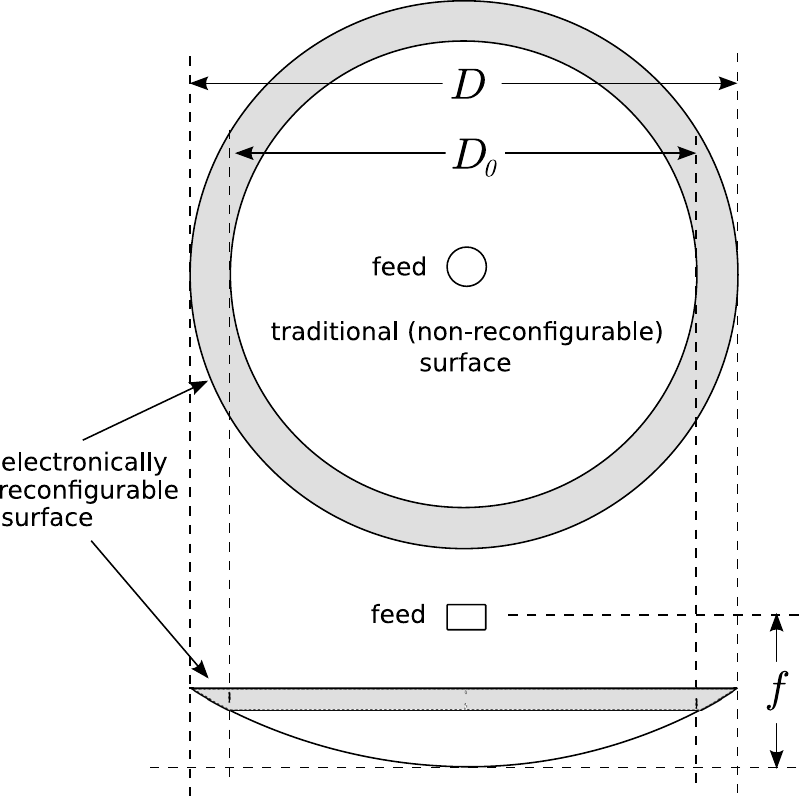}
\includegraphics[scale=0.5]{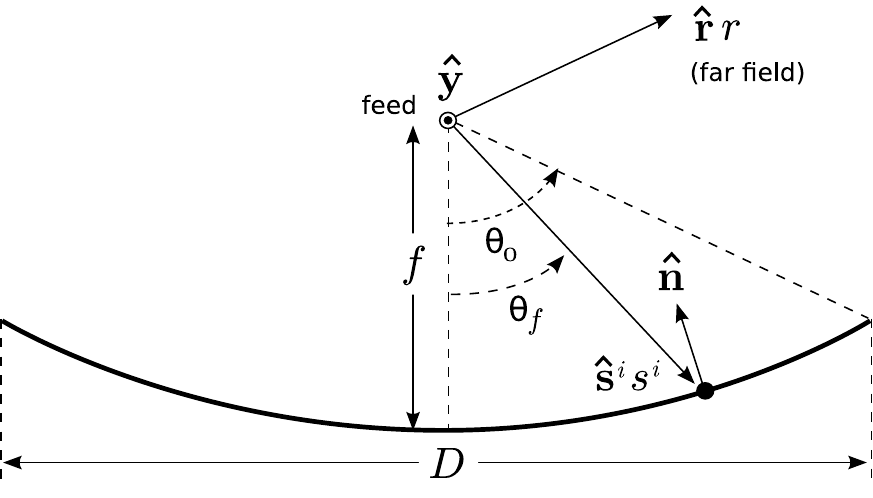}
\caption{On-axis ( top ) view of an electronically- reconfigurable rim scattering system assumed in this paper (reproduced from \cite{9400735}). The global origin is defined as the bottom of the dish. \(\theta_f\) is the angle measured from the reflector axis of rotation toward the rim with \(\theta_f = \theta_1\) at the rim of the fixed portion of the dish and \(\theta_f = \theta_0\) at the rim of the entire dish, \(\phi\) is the angular coordinate orthogonal to both \(\theta_f\) and the reflector axis. }
\label{fig:system_setting}
\end{center}
\end{figure}

We consider a prime focus axisymmetric circular paraboloidal reflector antenna system equipped with RIS elements along the outer rim, as illustrated in Fig.~\ref{fig:system_setting}. Following the development in~\cite{9400735}, we assume reciprocity between the transmit and receive patterns, and calculate the transmit patterns using physical optics (PO) \cite{stutzman2012antenna}. The total electric field intensity \(\mathbf{E}^s\) scattered by the reflector in the far-field direction with zenith angle \(\psi\) and azimuth angle \(\phi\) is given by 
\begin{equation}
\mathbf{E}^s(\psi, \phi) = \mathbf{E}_f^s(\psi, \phi) + \mathbf{E}_r^s(\psi, \phi),
\end{equation}
where \(\mathbf{E}_f^s\) represents the electric field intensity contributed from the fixed (non-reconfigurable) portion of the reflector (including any negligible feed-direct contribution), and \(\mathbf{E}_r^s\) accounts for the scattered field from the reconfigurable portion of the reflector. Specifically, \(\mathbf{E}_f^s\) can be written as
\begin{equation}
\mathbf{E}_f^s(\psi, \phi) = -j \omega \mu_0 \frac{e^{-j \beta r}}{4 \pi r} \int_{\theta_f = 0}^{\theta_1} \int_{\phi' = 0}^{2 \pi} \mathbf{J}_0 \left(\mathbf{s}^i \right) e^{j \beta \hat{\mathbf{r}} (\psi, \phi) \cdot \mathbf{r}'} \ \mathrm{d}s,
\label{eq:E_f_continue}
\end{equation}
where \(\omega\) is the angular frequency, \(\mu_0\) is the permeability of free space, \(\beta\) is the wavenumber, \(r\) denotes the distance between the origin and the far field observation point in the direction \(\psi\), and \(\mathrm{d}s\) is the differential element of surface area. The PO equivalent surface current distribution for the fixed portion of the dish \(\mathbf{J}_0 (\mathbf{s}^i) = 2 \hat{\mathbf{n}} \times \mathbf{H}^i(\mathbf{s}^i)\) is computed from the incident magnetic field \(\mathbf{H}^i(\mathbf{s}^i)\) and the surface normal \(\hat{\mathbf{n}}\). As in \cite{9400735}, we model the feed as
\begin{equation}
\mathbf{H}^i(\mathbf{s}^i) = I_0 \frac{\hat{\mathbf{y}} \times \hat{\mathbf{s}}^i}{\left| \hat{\mathbf{y}} \times \hat{\mathbf{s}}^i \right|} \frac{e^{-j \beta s^i}}{s^i} \cos(\theta_f)^q, 
\label{eq:mag_field}
\end{equation}
where \(I_0\) denotes the feed magnitude and phase, \(q\) controls the feed directivity, and \(\mathbf{s}^i = \hat{\mathbf{s}}^i s^i\) is the source-to-surface position vector with magnitude \(s^i\) pointing toward the direction \(\hat{\mathbf{s}}^i\) and can be written as
\begin{equation}
\mathbf{s}^i = \rho \cos(\phi')  \ \hat{\mathbf{x}} +  \ \rho \sin(\phi') \ \hat{\mathbf{y}} +  \left( \frac{\rho^2}{4f} -f \right) \hat{\mathbf{z}},
\end{equation}
where \(f\) represents the focal length, \(\rho  = 2f \tan(\theta_f/2)\) represents the radial distance on the \(xy\)-plane, and \(\phi'\) denotes the azimuth angle for integration. The magnitude \(\left|\mathbf{s}^i \right| =  s^i = \frac{\rho^2}{4f} + f\). The angle \(\theta_f\) is the angle measured from the reflector axis of rotation toward the rim with \(\theta_f = \theta_1\) at the rim of the fixed portion of the dish and \(\theta_f = \theta_0\) at the rim of the entire dish. \par
The position vector \(\mathbf{r}'\) points from the vertex of the dish (defined as the origin) to a point on the illuminated surface and can be expressed as
\begin{equation}
\mathbf{r}' = \rho \cos(\phi') \ \hat{\mathbf{x}} +  \rho \sin(\phi') \ \hat{\mathbf{y}} +  \frac{\rho^2}{4f} \ \hat{\mathbf{z}}. 
\end{equation}
The far field unit vector \(\hat{\mathbf{r}}(\psi, \phi)\) pointing towards the direction \((\psi, \phi)\) is 
\begin{equation}
\hat{\mathbf{r}}(\psi, \phi) = \sin(\psi)\cos(\phi) \ \hat{\mathbf{x}} + \sin(\psi)\sin(\phi) \ \hat{\mathbf{y}} +  \cos(\psi) \ \hat{\mathbf{z}}. 
\end{equation}
These geometric parameters are illustrated in Fig.~\ref{fig:system_setting}. The reflector geometry follows standard reflector antenna design principles in \cite{stutzman2012antenna}.  \par
In contrast to the fixed portion of the dish, the electric field intensity scattered by the reconfigurable portion of the reflector is similarly written as
\begin{equation}
\mathbf{E}_r^s(\psi, \phi) = -j \omega \mu_0 \frac{e^{-j \beta r}}{4 \pi r} \int_{\theta_f = \theta_1}^{\theta_0} \int_{\phi' = 0}^{2 \pi} \mathbf{J}_1 \left(\mathbf{s}^i \right) e^{j \beta \hat{\mathbf{r}} (\psi, \phi) \cdot \mathbf{r}'} \ \mathrm{d}s,
\end{equation}
where the major differences between the contributions due to the fixed and reconfigurable portions of the dish are (a) the angles in the integration and (b) the PO equivalent surface current distribution. Due to the discrete nature of the reconfigurable surface, we can write \(\mathbf{E}_r^s(\psi, \phi)\) as a summation over the contributions of individual elements, that is
\begin{equation}
\mathbf{E}_r^s (\psi, \phi) = -j \omega \mu_0 \frac{e^{-j \beta r}}{4 \pi r} \sum_n \mathbf{J}_1 \left(\mathbf{s}_n^i \right) e^{j \beta \hat{\mathbf{r}}(\psi, \phi) \cdot \mathbf{r}_n'} \Delta s,
\label{eq:E_discrete}
\end{equation}
where \(\mathbf{r}_n'\) can be parameterized by the radial distance \(\rho\) instead of the angle \(\theta_f\), and \(\Delta s\) denotes the area associated with each RIS element. \(\mathbf{J}_1 \left(\mathbf{s}_n^i \right) = w_n \mathbf{J}_0 \left(\mathbf{s}_n^i \right)\)\footnote{Here, the PO model treats each RIS element through an effective complex weight applied to its local scattered-field contribution. Therefore, finite-array mutual coupling and element-to-element interactions among differently configured RIS elements are not explicitly modeled. Such effects could be incorporated into the effective field matrix through full-wave simulation, measurement, or calibration; however, a detailed coupled-element electromagnetic design is beyond the scope of this work. } is the current distribution due to the \(n\)-th element with complex-valued weight \(w_n\). These weights will be designed to cancel sidelobes in specific directions of the co-polarized (co-pol) pattern. Specifically, we define the complex scalar \(E_f^{s, co} (\psi, \phi)\) and \(E_r^{s, co} (\psi, \phi)\) to represent the magnitude of \(\mathbf{E}_f^s (\psi, \phi)\) and \(\mathbf{E}_r^s (\psi, \phi)\) along the co-pol direction, respectively. The contribution of the \(N\) reconfigurable elements can be compactly written in the form:
\begin{equation}
E_r^{s, co} (\psi, \phi) = \mathbf{e}_{\psi, \phi}^T \mathbf{w},
\end{equation}
where \(\mathbf{w} \in \mathbb{C}^{N}\) denotes the complex RIS weights, and \(\mathbf{e}_{\psi, \phi} \in \mathbb{C}^N\) represents the co-pol portion of the electric field intensity without the influence of the RIS elements. The \(n\)-th element of \(\mathbf{e}_{\psi, \phi}\) is given by
\begin{equation}
e_{\psi, \phi, n} = \left(-j \omega \mu_0 \frac{e^{-j\beta r}}{4 \pi r} \mathbf{J}_0(\mathbf{s}_n^i) e^{j \beta \hat{\mathbf{r}}(\psi, \phi) \cdot \mathbf{r}_n'} \Delta s \right) \cdot \hat{\mathbf{e}}_{\mathrm{co}},
\label{eq:e_discrete}
\end{equation}
where the co-pol direction \(\hat{\mathbf{e}}_{\mathrm{co}}\) is defined as 
\begin{equation}
\begin{split}
\hat{\mathbf{e}}_{\mathrm{co}} &= \frac{\left( \hat{\mathbf{y}} \times \hat{\mathbf{r}} \right) \times \hat{\mathbf{r}}}{\left|  \left( \hat{\mathbf{y}} \times \hat{\mathbf{r}} \right) \times \hat{\mathbf{r}}\right|}, \\
&= \frac{1}{\sqrt{1 - \sin^2(\psi) \sin^2(\phi)}} \big( -\sin^2(\psi)\cos(\phi) \sin(\phi) \ \hat{\mathbf{x}} \\
+ &\left( \cos^2(\psi) + \sin^2(\psi) \cos^2(\phi) \right) \hat{\mathbf{y}} - \cos(\psi) \sin(\psi) \sin(\phi) \ \hat{\mathbf{z}} \big),
\label{eq:co_pol_direction}
\end{split}
\end{equation}
assuming that the feed is \(\hat{\mathbf{y}}\) polarized. 

Now to suppress the sidelobe gain at a specific direction, we wish
\begin{equation}
E_f^{s, co}(\psi, \phi) + E_r^{s, co}(\psi, \phi) = 0,
\end{equation}
which implies that the electric field intensity from the reconfigurable portion of the reflector must cancel that from the fixed portion of the reflector. Recalling that \(E_r^{s, co} (\psi, \phi) = \mathbf{e}_{\psi, \phi}^T \mathbf{w}\), the goal becomes finding the optimal weight vector \(\mathbf{w}^*\) that satisfies 
\begin{equation}
\mathbf{e}_{\psi, \phi}^T \mathbf{w}^* = -E_f^{s, co}(\psi, \phi).
\label{eq:p_single}
\end{equation}
Similarly, if we wish to place nulls in \(K\) distinct interference directions, which may correspond to \(K\) different interfering sources each operating at an arbitrary wavenumber (frequency of the interference), we can reformulate the problem (\ref{eq:p_single}) as
\begin{equation}
\mathbf{Aw}^* = \mathbf{y},
\label{eq:p_multi}
\end{equation}
where 
\begin{equation}
\begin{split}
\mathbf{A} &= \left[\mathbf{e}_{1}, \cdots, \mathbf{e}_{K} \right]^T \in \mathbb{C}^{K \times N}, \\
\mathbf{y} &= -\left[E_{f,1}^{s, co}, \dots, E_{f,K}^{s, co} \right]^T \in \mathbb{C}^{K \times 1}.
\end{split}
\end{equation}
Here, \(\mathbf{e}_i\) and \(E_{f, i}^{s, co}\) denote the co-pol electric field intensity vector and the electric field intensity from the fixed dish, respectively, corresponding to the direction \((\psi_i, \phi_i)\) at wavenumber (frequency) \(\beta_i\). \par
In many RIS-assisted nulling applications, it is desirable to suppress sidelobe interference while controlling the response in the desired mainlobe direction. Therefore, we include an optional mainlobe constraint at \(\psi=0^{\circ}\), given by \(\mathbf{e}_0^T\mathbf{w}^*=\kappa\), where \(\kappa\) specifies the desired contribution from the reconfigurable portion in the mainlobe direction. A natural choice is \(\kappa = \mathbf{e}_0^T \mathbf{1}\), which corresponds to preserving the original mainlobe contribution of the reconfigurable rim. However, with the weights restricted to unit modulus, imposing both the sidelobe nulling constraints and a strict mainlobe-preservation constraint reduces the size of the feasible set and potentially makes the resulting equality-constrained problem (\ref{eq:p_multi}) infeasible. Thus, \(\kappa\) can be selected according to the desired tradeoff between mainlobe control and sidelobe-nulling capability. For example, setting \(\kappa = 0\) forces the reconfigurable portion to make no contribution in the mainlobe direction, which may be suitable when the reconfigurable surface is intended only for interference suppression. In this work, we choose \(\kappa  = \delta E_{f,0}^{s, co}\), where \(0 < \delta \ll 1\). This choice uses the mainlobe field from the inner fixed-dish \(E_{f,0}^{s,\mathrm{co}}\) as the phase reference, so that the RIS contribution in the mainlobe direction is phase-aligned with the that of the fixed portion of the reflector. The scalar \(\delta\) controls the magnitude of this contribution, enabling a small constructive mainlobe contribution while retaining sufficient flexibility for sidelobe nulling.  Including the mainlobe constraint, the linear system in (\ref{eq:p_multi}) is augmented as follows: 
\begin{equation}
\begin{split}
\mathbf{A} &= \left[\mathbf{e}_{0}, \ \mathbf{e}_{1}, \cdots, \mathbf{e}_{K} \right]^T , \\
\mathbf{y} &= -\left[-\delta  E_{f,0}^{s, co}, \ E_{f,1}^{s, co}, \dots, E_{f,K}^{s, co} \right]^T,
\end{split}
\end{equation}
which provides a framework for jointly preserving the mainlobe gain and modifying the sidelobe pattern through optimization of the RIS weights. \par
Prior experimental demonstrations of programmable RIS/metasurface prototypes \cite{alexandropoulos2023ris, 10172310} support the practical feasibility of implementing reconfigurable reflecting elements. Therefore, the proposed architecture can be viewed as a conformal adaptation of programmable RIS elements to the reflector rim. In the reflector-antenna setting, the considered reconfigurable-rim model follows the system studied in \cite{9400735, 10403767}, where an L-band rim reflectarray for a \(D = 18\)m prime-focus reflector was designed and validated through CST full-wave simulation, and a \(1\)-bit reconfigurable unit-cell implementation was shown to enable sidelobe null steering. Moreover, since only a narrow rim portion is made reconfigurable and low-bit phase states are considered, the required hardware and control complexity is reduced compared with fully reconfigurable surfaces.

\section{Weight Selection}
\label{sec:III}
As discussed in \cite{10115669}, the approach to find the appropriate weights which satisfy (\ref{eq:p_multi}) depends on the restrictions placed on \(\mathbf{w}^*\). In this section, we present the methods for computing the optimal (or sub-optimal) weights under three regimes: (1) unconstrained, (2) continuous-phase unimodular, and (3) discrete-phase unimodular. 

\subsection{Unconstrained Weights}
If the weights are unconstrained, the linear system (\ref{eq:p_multi}) can be solved by using the pseudo inverse: 
\begin{equation}
\mathbf{w}^* = \mathbf{A}^{\dagger} \mathbf{y} = \mathbf{A}^H \left(\mathbf{AA}^H \right)^{-1} \mathbf{y}.
\end{equation}
We refer to \(\mathbf{w}^*\) as the unconstrained optimal weights, and we will demonstrate in the following section that it guarantees a zero response of the dish in the direction \((\psi, \phi)\). However, \(\mathbf{w}^*\) in general has elements with \(|w_n| \neq 1\) which requires that the elements have controllable gains (i.e., can provide attenuation or amplification to the scattered field). Such a requirement is undesirable from a cost and complexity perspective. Therefore, in most practical RIS implementations, we restrict the weights to satisfy \(|w_n| = 1\). 
\subsection{Unquantized Unimodular Weights}
When the weights are constrained to have unit magnitude (i.e., \(|w_n| = 1\)), solving (\ref{eq:p_multi}) is equivalent to solving the following minimization problem: 
\begin{equation}
\begin{split}
\min_{\mathbf{w}\in \mathbb{C}^{N}} \quad & f(\mathbf{w}) = \Vert \mathbf{Aw} - \mathbf{y}\Vert^2 \\
\mathrm{s. t. } \; \quad  &|w_n| = 1, \quad n = 1, \dots, N,
\label{eq:LS}
\end{split}
\end{equation}
This is a nonconvex, complex-valued, constant-modulus, least squares optimization problem which is a special case of nonconvex quadratically-constrained quadratic programming. Denote \(\mathbf{w}_c^*\) as the stationary point of the problem (\ref{eq:LS}). One solution to this problem is to use semi-definite relaxation \cite{5447068}. Unfortunately, this expands the problem such that it is of dimension \(N^2\). Given that for moderate to large antennas (\(D = 10\)m to 100m) or high operating frequencies, the number of RIS elements \(N\) can range from \(10^3\) to \(10^5\), such an expansion is computationally infeasible. Alternatively, problem (\ref{eq:LS}) can be addressed using first-order optimization methods, for which empirical results show that the global minimum can be obtained in most scenarios. We describe two such approaches below.
\subsubsection{Gradient Projection}
Gradient projection iteratively updates the solution using a gradient descent step followed by a projection onto the set of unit-modulus vectors. Although the unit-modulus constraint is nonconvex, the algorithm has been proven to converge to stationary points \cite{7849224, 10284537}. The details of GP are summarized in Algorithm~\ref{alg:GP}. In the algorithm, \(\angle(\cdot) \) denotes the phase of the element inside.
\begin{algorithm}[H]
\caption{Gradient Projection.}
\begin{algorithmic}
\STATE 
\STATE Initialization: Set \(k = 0\), \(\alpha = \frac{1}{\lambda_1 (\mathbf{AA}^H)}\), \(\mathbf{w}^{(0)} = \mathbf{1}\).
\STATE \textbf{Repeat: }
\STATE \hspace{0.5cm} \(\boldsymbol{\eta}^{(k+1)} = \mathbf{w}^{k} - \alpha \mathbf{A}^H \left(\mathbf{Aw}^{(k)} - \mathbf{y} \right)\);
\STATE \hspace{0.5cm} \(\mathbf{w}^{(k+1)} = e^{j\angle \left(\boldsymbol{\eta}^{(k+1)} \right)}\);
\STATE \hspace{0.5cm} \(k = k + 1\);
\STATE \textbf{Until Convergence}
\end{algorithmic}
\label{alg:GP}
\end{algorithm}

\subsubsection{Alternating Projection}
Another effective method is alternating projection, which views the problem (\ref{eq:LS}) as finding a point in the intersection of two sets, or a point that yields the minimum distance between them, where the two sets are:
\begin{equation}
\begin{split}
\mathcal{S}_1 = &\left\{\mathbf{w} \in \mathbb{C}^N \ \big| \ \mathbf{Aw} = \mathbf{y} \right\}, \\
\mathcal{S}_2 = &\left\{\mathbf{w} \in \mathbb{C}^N \ \big| \ |w_n| = 1, n = 1, \dots, N \right\}. 
\end{split}
\end{equation}
The alternating projection algorithm starts with an initial point and then alternatively projects the point onto \(\mathcal{S}_1\) and \(\mathcal{S}_2\). The projection operation onto \(\mathcal{S}_1\) is equivalent to solving the problem \(\Pi_{\mathcal{S}_1}(\mathbf{w})=\)
\begin{equation}
\begin{split}
\min_{\mathbf{w}\in \mathbb{C}^{N}} \quad & \Vert \mathbf{w} - \mathbf{x}\Vert_2^2 \\
\mathrm{s. t. } \; \quad  & \mathbf{Ax} = \mathbf{y},
\end{split}
\end{equation}
and the solution can be written as \(\Pi_{\mathcal{S}_1}(\mathbf{w})= \mathbf{w} - \mathbf{A}^{\dagger}\left(  \mathbf{Aw} - \mathbf{y}\right)\). On the other hand, the projection onto \(\mathcal{S}_2\) is again given by \(\Pi_{\mathcal{S}_2}(\mathbf{w})= e^{j \angle(\mathbf{w})}\). The implementation details of alternating projection are summarized in Algorithm~\ref{alg:AP}.
\begin{algorithm}[H]
\caption{Alternating Projection.}
\begin{algorithmic}
\STATE 
\STATE Initialization: Set \(k = 0\), \(\mathbf{w}^{(0)} = \mathbf{1}\).
\STATE \textbf{Repeat: }
\STATE \hspace{0.5cm} \(\boldsymbol{\eta}^{(k+1)} = \mathbf{w}^{k} - \mathbf{A}^{\dagger} \left(\mathbf{Aw}^{(k)} - \mathbf{y} \right)\);
\STATE \hspace{0.5cm} \(\mathbf{w}^{(k+1)} = e^{j \angle \left(\boldsymbol{\eta}^{(k+1)} \right)}\);
\STATE \hspace{0.5cm} \(k = k + 1\);
\STATE \textbf{Until Convergence}
\end{algorithmic}
\label{alg:AP}
\end{algorithm}
We notice that the only difference between GP and AP is the gradient descent step or the projection step \(\Pi_{\mathcal{S}_1}(\mathbf{w})\). And this difference offers an advantage to AP in that it converges much faster than GP, and it does not require the initialization of step size. While the convergence for the alternating projection is well established when two sets are both convex sets, it is not guaranteed when one of the sets is nonconvex (\(\mathcal{S}_2\)). In \cite{9681803}, the authors show local convergence of AP when the two sets have a nonempty intersection and the initialization is sufficiently close to an intersection point. They also discuss additional conditions under which convergence from arbitrary initialization can be established; in particular, the entries of \(\Pi_{\mathcal{S}_1}\left(\mathbf{w}^{(k)}\right)\) must be bounded away from zero. Although this condition cannot be verified analytically for our system setting, our empirical results show that AP typically reaches a zero-residual solution when only a small number of nulling angles are imposed. In these cases, the achieved null depth is around \(-300\) dBi, which corresponds to numerical satisfaction of \(\mathbf{A}\mathbf{w}=\mathbf{y}\). \par
On the other hand, if \(\mathcal{S}_1\) and \(\mathcal{S}_2\) do not intersect, there is no unimodular weight vector that exactly satisfies the linear equality \(\mathbf{A}\mathbf{w}=\mathbf{y}\), and perfect nulling under the unimodular constraint is therefore infeasible. Consequently, AP may converge to a fixed point of the alternating-projection iteration, corresponding to a pair of nearby points, one in the affine constraint set \(\mathcal{S}_1\) and one in the unit-modulus set \(\mathcal{S}_2\). However, such convergence is not guaranteed by the local convergence theorem discussed in \cite{9681803}. Moreover, because \(\mathcal{S}_2\) is nonconvex, the resulting fixed point is not guaranteed to be the globally closest pair between the two sets, nor is the resulting unimodular vector guaranteed to minimize \(\Vert\mathbf{A}\mathbf{w}-\mathbf{y}\Vert^2\). This explains why AP may yield a larger nonzero residual than GP. As demonstrated in Section~\ref{sec:V}, each method therefore offers distinct advantages depending on the scenario. \par

\subsection{Simplified Unquantized Unimodular Weights} \label{sec:apprx}
Any two complex numbers whose magnitudes sum to a value less than 2 can be replaced by two unimodular complex numbers. Additionally, in our problem, the two neighboring elements in each \(\mathbf{e}_i\) are nearly identical. Therefore, in the cases where \(||\mathbf{w}^*||_{\infty} \leq 1\), it is guaranteed that we can replace the optimal weights \(\mathbf{w}^*\) with approximately optimal unimodular weights \(\tilde{\mathbf{w}}_c^*\) which satisfy \(\mathbf{A}\tilde{\mathbf{w}}_c^* \approx \mathbf{Aw}^*\). To do this, first we group every two adjacent columns of \(\mathbf{A}\) into one column:
\begin{equation}
\tilde{\mathbf{A}}_n = (\mathbf{A}_{2n-1} + \mathbf{A}_{2n})/2 , \quad n = 1, \dots, \frac{N}{2}, 
\end{equation}
where \(\mathbf{A}_n\) represents the \(n\)-th column of matrix \(\mathbf{A}\). Now, we require that
\begin{equation}
\begin{split}
\sum_{n=1}^{\frac{N}{2}} \tilde{\mathbf{A}}_n (\tilde{w}_{c,2n - 1}^* + \tilde{w}_{c,2n}^*) & \approx \sum_{n=1}^N \mathbf{A}_n w_n^*,
\end{split}
\end{equation}
which means that for each pair of weight elements, we need 
\begin{equation}
\tilde{\mathbf{A}}_n (\tilde{w}_{c,2n - 1}^* + \tilde{w}_{c,2n}^*) \approx \mathbf{A}_{2n-1} w_{2n - 1}^* + \mathbf{A}_{2n}w_{2n}^*. 
\label{eq:approx_pair}
\end{equation}
By defining \(\tilde{w}_{2n-1} = e^{j \theta_{n,1}}\), \(\tilde{w}_{2n} = e^{j \theta_{n, 2}}\), and assuming the equality can be met in (\ref{eq:approx_pair}), (\ref{eq:approx_pair}) can be rewritten as
\begin{equation}
\begin{split}
\cos(\theta_{n, 1}) + \cos(\theta_{n, 2}) &= \Re(y_n),\\
\sin(\theta_{n, 1}) + \sin(\theta_{n, 2}) &= \Im(y_n),\\
\end{split}
\end{equation}
where \(y_n = (\mathbf{A}_{2n-1} w_{2n - 1}^* + \mathbf{A}_{2n}w_{2n}^*) \oslash \tilde{\mathbf{A}}_n\) and \(\oslash\) denotes the Hadamard (element wise) division. Solving the above equation, we get
\begin{equation}
\begin{split}
\theta_{{n, 1}} &= \angle (y_n) + \cos^{-1} \left( \frac{|y_n|}{2} \right), \\
\theta_{{n, 2}} &= \angle (y_n) - \cos^{-1} \left( \frac{|y_n|}{2} \right).
\end{split}
\end{equation}
Therefore, each pair of the weight elements can be written as 
\begin{equation}
\tilde{\mathbf{w}}_c^* = \left[ e^{j\theta_{1,1}}, \ e^{j\theta_{1,2}},  \cdots,  e^{j \theta_{N/2, 1}}, \ e^{j \theta_{N/2, 2}} \right]^T.
\end{equation}
This simplified unimodular solution provides acceptable performance when compared to the optimal continuous unimodular weights \(\mathbf{w}_c^*\), as we will see in Section~\ref{sec:V}. Importantly, since \(\tilde{\mathbf{w}}_c^*\) can be written in closed form as a function of the unconstrained optimal weights \(\mathbf{w}^*\), it requires less computation time than any of the iterative algorithms discussed.

\subsection{Unimodular Quantized Weights}
The previous approaches assume that RIS elements support continuous-phase values. However, in practical implementations, each element is typically constrained to a finite set of discrete phase shifts due to hardware limitations. This motivates solving the problem (\ref{eq:LS}) with a new constraint such that
\begin{equation}
\begin{split}
\min_{\mathbf{w}\in \mathbb{C}^{N}} \quad & f(\mathbf{w}) = \Vert \mathbf{Aw} - \mathbf{y}\Vert^2 \\
\text{s.t. } \; \quad  &w_n \in \mathcal{W}, \quad n = 1, \dots, N, 
\label{eq:LS2}
\end{split}
\end{equation}
where \(\mathcal{W} = \left\{w \in \mathbb{C} \big| w = e^{j \frac{2 \pi k}{M}}, k = 0, \dots, M-1 \right\}\) and \(M\) is the number of quantization levels. In our previous work \cite{10115669}, we addressed this problem using a serial search algorithm for binary weights, which sequentially flips each element of the weight vector and retains the change only if the new weight vector decreases the objective function. We also employed simulated annealing (SA) algorithm to find weights with higher quantization levels, in which the phase of one weight element is randomly changed at each step; the algorithm accepts the new weight vector if the cost is lower or accepts the new weight vector with a certain probability if the cost is higher. Although the implementation is straightforward, these metaheuristic approaches have the disadvantage that their convergence rate and optimality are not guaranteed. Moreover, the lack of consistency can be problematic in applications where reproducibility is important. \par
A more efficient solution is to use the negative square penalty (NSP) or the extreme point pursuit (EXPP) method recently proposed in \cite{8811616}. Specifically, rewriting problem (\ref{eq:LS2}) as
\begin{equation}
\min_{\mathbf{w} \in \bar{\mathcal{W}}^N} \quad F_c(\mathbf{w}) = \Vert \mathbf{Aw} - \mathbf{y}\Vert^2 - c \Vert\mathbf{w}\Vert^2,
\label{eq:NSP}
\end{equation}
where \(\bar{\mathcal{W}} = \text{conv}(\mathcal{W})\) is the convex hull of \(\mathcal{W}\). The penalty term \(-c \Vert \mathbf{w} \Vert_2^2\) forces each element of \(\mathbf{w}\) towards the extreme points of \(\bar{\mathcal{W}}\), which is \(\mathcal{W}\) itself. The authors in \cite{9681803} proved that when the objective function \(f\) has a \(L\)-Lipschitz continuous gradient on a convex set, problem (\ref{eq:NSP}) and problem (\ref{eq:LS2}) share the same optimal solution sets for any \(c > \frac{L}{2}\). Since our objective function is a quadratic function, \(L\) is equal to the maximum eigenvalue of \(\mathbf{A}^H \mathbf{A}\). \par
While the feasible set \(\bar{\mathcal{W}}^N\) is now convex, the objective function of (\ref{eq:NSP}) is no longer convex. An optimization problem with a non-convex objective function and a convex constraint can be solved using algorithms that break the original problem into manageable sub-problems. For a quadratic objective function, the Majorize-Minimization (MM) algorithm is preferable since it has a strong monotonicity guarantee and is more flexible in designing the surrogate function than other algorithms such as sequential convex approximation (SCA) or sequential convex programming (SCP) algorithms. Specifically, the MM algorithm is an iterative method which constructs and minimizes a series of surrogate functions of the objective function \(F_c\). The surrogate function \(g\left(\mathbf{w}|\mathbf{w}^{(k)} \right)\), which is also called a majorant at the \(k\)-th step, must satisfy
\begin{equation}
\begin{split}
g\left( \mathbf{w} \bigl\rvert \mathbf{w}^{(k)} \right) &\geq F_c(\mathbf{w}), \\
g\left( \mathbf{w}^{(k)} \bigl\rvert \mathbf{w}^{(k)} \right) &= F_c\left(\mathbf{w}^{(k)} \right), \quad \forall \ \mathbf{w}.
\end{split}
\end{equation}
These properties ensure that optimizing the majorant will either minimize the value of \(F_c\) or leave it unchanged, therefore guaranteeing monotonic descent in the objective function at each iteration. Using the fact that 
\begin{equation}
\begin{split}
&\bigm\Vert\mathbf{w} - \mathbf{w}^{(k)} \bigm\Vert^2 \geq 0, \qquad \text{for all }\mathbf{w}, \mathbf{w}^{(k)} \\
\Rightarrow & -\Vert \mathbf{w} \Vert^2 \leq \bigm\Vert \mathbf{w}^{(k)} \bigm\Vert^2 - 2 \left( \mathbf{w}^{(k)} \right)^H\mathbf{w},\\
& \qquad \quad \ \: = -2 \left( \mathbf{w}^{(k)} \right)^H \left( \mathbf{w} - \mathbf{w}^{(k)} \right) - \bigm\Vert \mathbf{w}^{(k)} \bigm\Vert^2,
\end{split}
\end{equation}
the majorant for \(F_c\) can be expressed as
\begin{equation}
\begin{split}
F_c(\mathbf{w}) &\leq f(\mathbf{w}) - 2 c  \left( \mathbf{w}^{(k)} \right)^H \left( \mathbf{w} - \mathbf{w}^{(k)} \right) - c \bigm\Vert \mathbf{w}^{(k)} \bigm\Vert^2,\\
& = g_c\left( \mathbf{w} \bigl\rvert \mathbf{w}^{(k)} \right),
\end{split}
\end{equation}
and we have 
\begin{equation}
F_c\left(\mathbf{w}^{(k)} \right) = g\left( \mathbf{w}^{(k)} \bigl\rvert \mathbf{w}^{(k)} \right),
\end{equation}
which satisfies the properties of a majorant. Consequently, the solution at the (\(k+1\))-th step becomes
\begin{equation}
\mathbf{w}^{(k+1)} = \argmin_{\mathbf{w} \in \Bar{\mathcal{W}}^N} \ g\left(\mathbf{w} \bigl\rvert \mathbf{w}^{(k)} \right), \quad k = 0, 1, \dots.
\label{eq:surr}
\end{equation}

Although each sub-problem (\ref{eq:surr}) is convex, solving it in exact fashion at every iteration is computationally expensive. To improve efficiency, we adopt the one-step accelerated gradient projection (APG) strategy proposed in \cite{8811616}, which finds an approximate solution for the problem (\ref{eq:surr}) and proceeds to the next update within each step. To further reduce the computational time, we omit the backtracking line search step which is used to find the step size at each step \(k\). Specifically, at the \(k\)-th step, we generate an extrapolated point \(\mathbf{z}^{(k)}\) which carries the ``momentum" from \(\mathbf{w}^{(k)}\) and \(\mathbf{w}^{(k-1)}\), then perform gradient descent only once at \(\mathbf{z}^{(k)}\) over the majorant \(g_c \left(\mathbf{z}^{(k)} \bigl\rvert \mathbf{w}^{(k)} \right)\) to find \(\mathbf{z}^{(k+1)}\). Finally, we obtain \(\mathbf{w}^{(k+1)}\) by projecting \(\mathbf{z}^{(k+1)}\) back to the convex set \(\Bar{\mathcal{W}}^N\). The projection operation for the constraint of discrete phase is more complicated than the case of continuous phase. Instead of solving the optimization problem of projection, the authors in \cite{8811616} proposed a closed form expression of the projection: 
\begin{equation}
\Pi_{\Bar{\mathcal{W}}^N}(w) = e^{j \frac{2 \pi m}{M}} \left( [\Re(\bar{w})]_0^{\cos(\pi/M)} + j[\Im(\bar{w})]_{-\sin(\pi/M)}^{\sin(\pi/M)} \right),
\end{equation}
where 
\begin{equation}
m = \biggm\lfloor \frac{\angle (w) + \pi/M}{2\pi/M} \biggm\rfloor, \quad \bar{w} = w e^{-j \frac{2\pi m}{M}},
\end{equation}
and \([\mathbf{w}]_\mathbf{a}^\mathbf{b}\) denotes that for each element \(w_n\) in \(\mathbf{w}\),
\begin{equation}
w_n = \min\{b_n, \max\{w_n, a_n\}\}. 
\end{equation}

Based on our empirical results, we found that the algorithm is more likely to be stuck in a poor local minimum if we choose a large penalty parameter \(c\) at the beginning. Therefore, given that our original objective function \(f(\mathbf{w})\) is convex, we start with \(c = 0\) and gradually increase the value to \(c_{\max} > \lambda_1 \left(\mathbf{A}^H \mathbf{A} \right) / 2\). It turns out that this strategy provides good convergence while guaranteeing that the solution will eventually lie on an extreme point which is a feasible solution for the original problem (\ref{eq:LS2}). The implementation details are provided in Algorithm~\ref{alg:EXPP}.
\begin{algorithm}[H]
\caption{Extreme Point Pursuit (EXPP).}
\begin{algorithmic}
\STATE 
\STATE Initialization: Set \(k = 0\), \(\mathbf{w}^{(-1)} = \mathbf{w}^{(0)} = \mathbf{1}\), \(\xi_{-1} = 0\), \(c = 0\) Choose \(\beta > \lambda_1 \left(\mathbf{A}^H\mathbf{A} \right)\) and \(c_{\max} > \frac{\lambda_1 \left(\mathbf{A}^H \mathbf{A} \right)}{2}\).
\STATE \textbf{Repeat: }
\STATE \hspace{0.5cm} \(\xi_k = \frac{1 + \sqrt{1 + 4 \xi_{k-1}^2}}{2}\), \(\alpha_k = \frac{\xi_{k-1} - 1}{\xi_k}\);
\STATE \hspace{0.5cm} \(\mathbf{z}^{(k)} = \mathbf{w}^{(k)} + \alpha_k \left(\mathbf{w}^{(k)} - \mathbf{w}^{(k-1)} \right)\);
\STATE \hspace{0.5cm} \(\mathbf{w}^{(k+1)} = \Pi_{\bar{\mathcal{W}}^N} \left( \mathbf{z}^{(k)} - \frac{1}{\beta} \nabla g_c \left(\mathbf{z}^{(k)} \bigl\rvert \mathbf{w}^{(k)} \right) \right)\);
\STATE \hspace{0.5cm} linearly increase the value of \(c\) towards \(c_{\max}\), \(k = k + 1\);
\STATE \textbf{Until Convergence}
\end{algorithmic}
\label{alg:EXPP}
\end{algorithm}\par
Simulation results in Section~\ref{sec:V} demonstrate that the EXPP algorithm not only exhibits a faster convergence rate compared to that of the simulated annealing, but also converges to better stationary points, especially as the number of angles to null increases. \par
We note that practical RIS elements may exhibit phase-dependent amplitude variations, rather than maintaining exactly unit amplitude for all phase states \cite{9148961}. The unit-modulus constraint considered here provide useful baseline cases for evaluating the achievable nulling performance under ideal phase control. If a practical phase-shift model is available, the proposed formulation can be modified by replacing the unit-modulus feasible set with a hardware-dependent feasible set. For example, instead of \(w_n = e^{j\theta_n}\), the element response can be written as \(w_n = a(\theta_n)e^{j\theta_n}\), where \(a(\theta_n)\) captures the phase-dependent amplitude response. In this case, the resulting problem remains a constrained LS problem, but the feasible set is changed to a hardware-characterized response curve or points. Consequently, optimization methods based on projected updates can still be applied if the projection onto this feasible set is available. However, the convergence analysis would need to be performed on a case-by-case basis, since it depends on the geometry of the modified feasible set, which is determined by the specific hardware response model. \par
We also note that deep network-based methods could be considered for the nonconvex optimization problems studied in this section. For the GP and AP algorithms considered above, global optimality is satisfied when the algorithm reaches a zero-residual solution, i.e., when \(\mathbf{A}\mathbf{w}=\mathbf{y}\) is numerically satisfied. In this case, the LS objective attains its global lower bound of zero, and therefore no further improvement in the nulling objective is possible. This situation typically occurs when nulling a small number of angles. In such cases, a neural network is not needed to improve the final nulling objective, although it may still be useful for reducing online computation or providing fast initialization. In more challenging cases where GP or AP does not reach a zero-residual solution, the nonconvexity of the feasible set prevents a global optimality guarantee. In such cases, learning-based methods may provide improvement through better initialization and acceleration. However, a systematic comparison with deep network-based methods would require problem-specific network design, training-data generation, and generalization testing across different nulling scenarios and hardware constraints, which is left for future work. 

\subsection{Computational Complexity Analysis}
The computational complexity of the proposed channel estimation method is analyzed in this subsection. Given that \(\mathbf{A} \in \mathbb{C}^{(K+1) \times N}\), \(\mathbf{y} \in \mathbb{C}^{K+1}\), and \(K \ll N\), the dominant cost in each GP iteration in Algorithm~\ref{alg:GP} comes from the two matrix-vector multiplications \(\mathbf{A}\mathbf{w}^{(k)}\) and \(\mathbf{A}^H\left (\mathbf{A}\mathbf{w}^{(k)}-\mathbf{y} \right)\). Therefore, the complexity of GP with \(T_{\mathrm{GP}}\) iterations is \(\mathcal{O}(T_{\mathrm{GP}}(K+1)N)\), excluding the one-time computation of the step size. Similarly, for the AP algorithm in Algorithm~\ref{alg:AP}, the total iterative complexity of AP algorithm is \(\mathcal{O}(T_{\mathrm{AP}}(K+1)N)\), given that \(\mathbf{A}^{\dagger}\) can be precomputed.  \par
For the discrete-phase unimodular constraint, the simulated annealing algorithm with \(T_{\mathrm{SA}}\) iterations has the complexity \(\mathcal{O}(T_{\mathrm{SA}}(K+1)N)\) \cite{russell1995modern}, since each iteration evaluates the objective function by computing the gain associated with the current weight vector. For the EXPP algorithm, the computational complexity is dominated by the gradient calculation \(\nabla g_c\), while the projection \(\Pi_{\bar{\mathcal{W}}^N}\) is performed element-wise in closed form. Since the gradient calculation involves matrix-vector multiplications with \(\mathbf{A}\) and \(\mathbf{A}^H\), the total complexity of EXPP with \(T_{\mathrm{EXPP}}\) iterations is \(\mathcal{O}(T_{\mathrm{EXPP}}(K+1)N)\). All iterative algorithms considered here have comparable per-iteration complexity, and their practical computational costs are mainly determined by the number of iterations required for convergence. 

\section{Properties of the Electric Field Intensity Vectors}
\label{sec:IV}
While the unconstrained problem can always be solved in closed form, the introduction of the unimodular constraint can make the system infeasible, especially as the number of angles to null increases. To better understand this limitation, we explore the structure and correlation characteristics of \(\mathbf{e}_i\) that make up the matrix \(\mathbf{A}\). 

\subsection{Correlation of the Electric Field Intensity Vector for Different Angles and Frequencies}
\label{sec:IV-A}
\begin{figure*}[h]
\centering
\begin{subfigure}{.31\linewidth}
  \centering
  \includegraphics[width=1.05\linewidth]{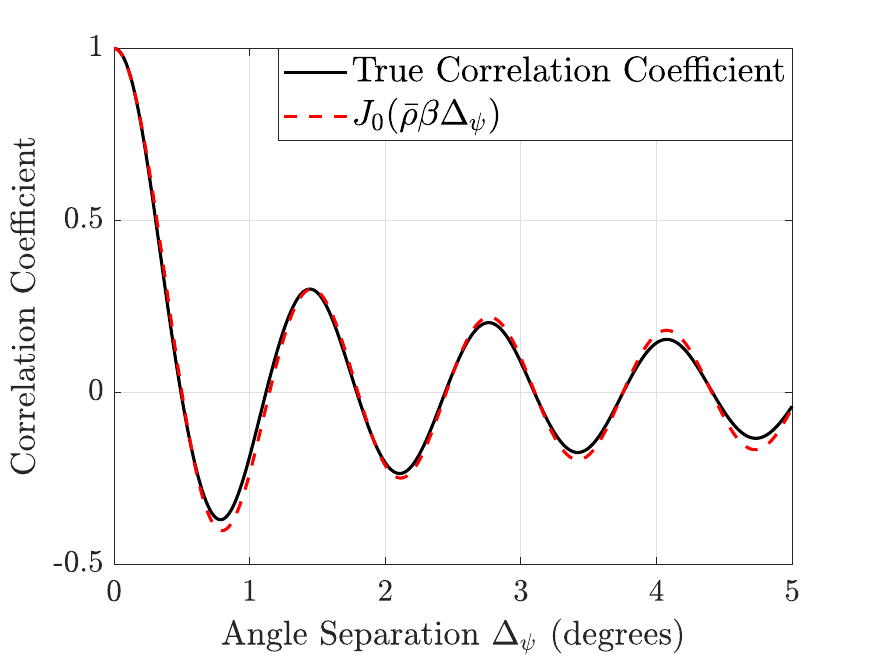}
\end{subfigure}%
\begin{subfigure}{.31\textwidth}
  \centering
  \includegraphics[width=1.05\linewidth]{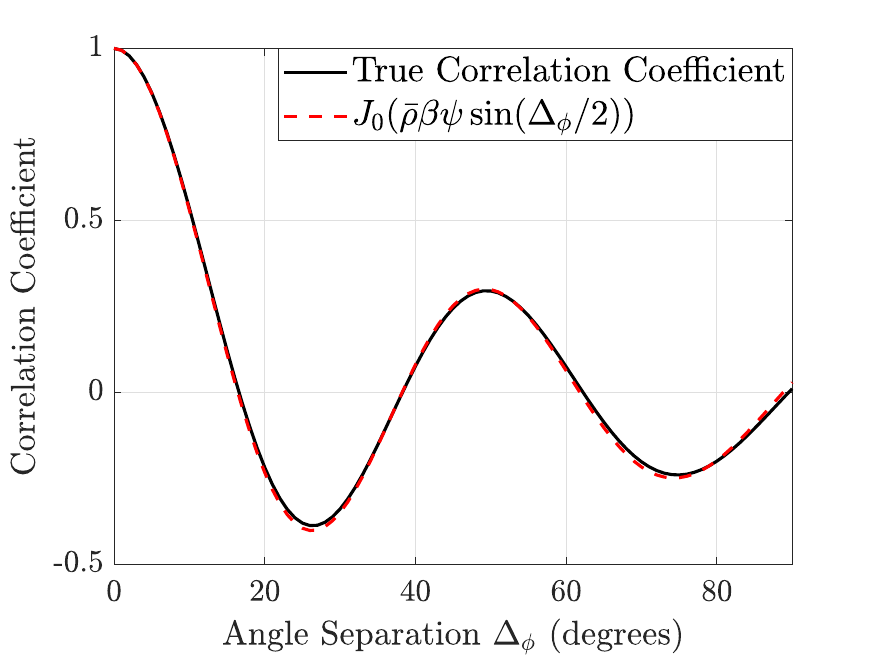}
\end{subfigure}%
\begin{subfigure}{.31\textwidth}
  \centering
  \includegraphics[width=1.05\linewidth]{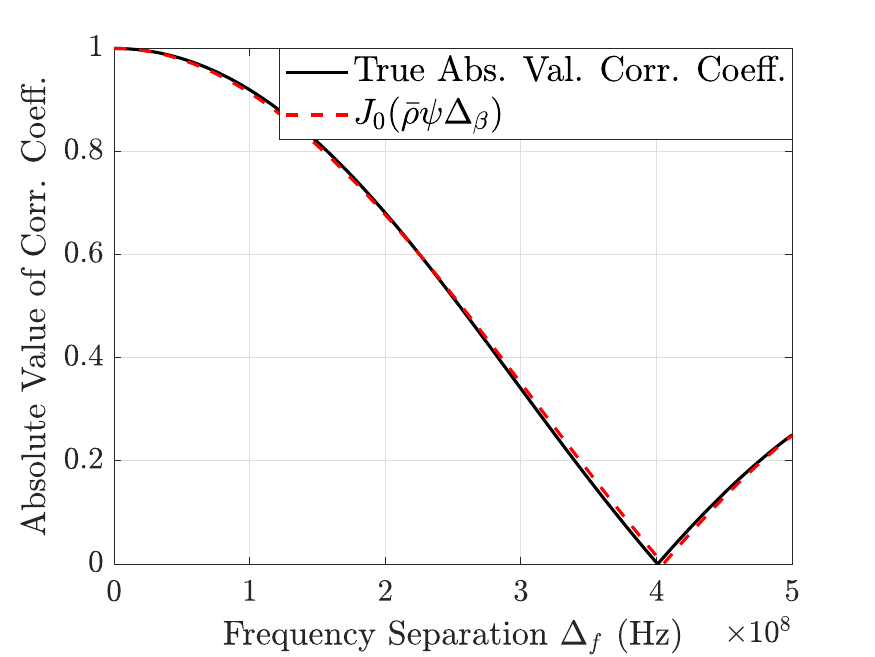}
\end{subfigure}
\caption{Correlation coefficient versus the zenith angle separation \(\Delta_\psi\) (left) and the azimuth angle separation \(\Delta_\phi\) (middle). Absolute value of the correlation coefficient versus frequency separation \(\Delta_f\) (right). }
\label{fig:J_0}
\end{figure*}

Since the entries of \(\mathbf{e}_i\) are very small, direct comparison of the correlations between the vectors \(\mathbf{e}_i\) does not clearly reveal the differences. Instead, we analyze the correlation coefficient, which highlights the relative similarity independent of the scale. Defining \(c_n = e_{2, n}/e_{1, n}\) as the ratio between the \(n\)-th elements at two electric field intensity vectors for any two given far field directions \((\psi_1, \phi_1)\) and \((\psi_2, \phi_2)\) with respective wavenumbers (frequencies) \(\beta_1\) and \(\beta_2\), the correlation coefficient \(C\) between \(\mathbf{e}_{1}\) and \(\mathbf{e}_{2}\) is approximated by
\begin{equation}
\begin{split}
C(\mathbf{e}_{1}, \mathbf{e}_{2}) &= \frac{1}{N-1} \sum_{n = 1}^N \left( \frac{e_{1, n} - \mu_{1}}{\sigma_{1}} \right)^* \left( \frac{e_{2, n} - \mu_{2}}{\sigma_{2}} \right),\\
&\approx \frac{1}{N-1} \sum_{n = 1}^N \frac{|e_{1, n}|^2 c_n}{\sigma^2} \approx \frac{1}{N} \sum_{n = 1}^N  c_n,
\label{eq:corr_coef_def}
\end{split}
\end{equation}
where \(\mu_i\) and \(\sigma_i\) are the mean and standard deviation of the vector \(\mathbf{e}_{i}\), respectively. In the approximations above, we assume that (1) \(\mu_{i} \approx 0\) due to the central limit theorem since the number of elements \(N\) is large, and (2) \(\sigma_{1} \approx \sigma_{2} = \sigma\) based on the empirical results, where \(\sigma^2 \approx \mathbf{E}\left[ |e_{i,n}|^2 \right]\). Using (\ref{eq:mag_field}) and (\ref{eq:e_discrete}), we can approximate \(c_n\) as (see Appendix~\ref{app:A})
\begin{equation}
\begin{split}
c_n \approx e^{j \rho_n \left( \cos\left( \phi_n' \right)a  + \sin \left(\phi_n' \right) b \right) } e^{- j \Delta_\beta f},
\end{split}
\end{equation}
where \(\Delta_\beta = \beta_2 - \beta_1\), \(\rho_n\) and \(\phi_n'\) are the radial distance and azimuth angle of the \(n\)-th element, respectively. The correlation coefficient can be approximated as (see Appendix~\ref{app:A})
\begin{equation}
C(\mathbf{e}_{1}, \mathbf{e}_{2}) \approx J_0 \left(\bar{\rho} \sqrt{a^2 + b^2} \right) e^{-j \Delta_\beta f},
\label{eq:corr_coef_analy}
\end{equation}
where \(J_0(\cdot)\) is the zeroth-order Bessel function of the first kind, \(\bar{\rho}\) is the average radial distance of the reconfigurable rim, \(a = \beta_2 \psi_2 \cos(\phi_2) - \beta_1 \psi_1 \cos(\phi_1)\), and \(b = \beta_2 \psi_2 \sin(\phi_2) - \beta_1 \psi_1 \sin(\phi_1)\). We also note from (\ref{eq:corr_coef_analy}) that 
\begin{equation}
C(\mathbf{e}_{1}, \mathbf{e}_{2}) = 
\begin{cases}
J_0(\bar{\rho} \beta \Delta_{\psi} ), \qquad \qquad &\text{if } \beta_1 = \beta_2 = \beta, \\ &\ \ \ \phi_1 = \phi_2 = \phi, \\
J_0(2\bar{\rho} \beta \psi \sin(\Delta_\phi/2)) , \quad &\text{if } \psi_1 = \psi_2 = \psi \\ &\ \ \  \beta_1 = \beta_2 = \beta, \\
J_0(\bar{\rho} \psi \Delta_\beta) e^{-j \Delta_\beta f}, \quad &\text{if } \psi_1 = \psi_2 = \psi \\ &\ \ \ \phi_1 = \phi_2 = \phi, \\
\end{cases}
\label{eq:corr_coef_cases}
\end{equation}
where \(\Delta_{\psi} = \psi_2 - \psi_1\), and \(\Delta_{\phi} = \phi_2 - \phi_1\). Fig.~\ref{fig:J_0}  illustrates how the correlation coefficient behaves in each case. From (\ref{eq:corr_coef_cases}), we see that the electric field intensity vectors corresponding to different angles and frequencies are not co-linear, as their correlation is strictly between \(-1\) and 1. Therefore, for \(K\) different directions and frequencies, the matrix \(\mathbf{A} \in \mathbb{C}^{K \times N}\) has full rank and its column space spans the entire \(K\)-dimensional space. This guarantees that any vector \(\mathbf{y}\) lies within the column space of \(\mathbf{A}\), and for sufficiently large \(N\), there always exists an unconstrained solution \(\mathbf{w}^*\) that strictly satisfies the linear equality, i.e., creates perfect nulls for all desired directions and frequencies. \par

For modern communication systems, the carrier frequencies of the communication signals are typically in the MHz to GHz range, which leads to large corresponding wavernumbers. On the other hand, since the antenna pattern usually has a small gain beyond the first few sidelobes, the zenith angle of interest is small. Therefore, given that \(\beta \gg \psi\), the first case of (\ref{eq:corr_coef_cases}) indicates that small perturbations in \(\psi\) result in significant changes in the argument of the Bessel function, and thus in the correlation coefficient between the electric field intensity vectors. In contrast, for a fixed zenith angle \(\psi\), the correlation decays more slowly with respect to the frequency separation \(\Delta_\beta\) and the azimuth angle separation \(\Delta_\phi\). Therefore, the correlation between the electric field intensity vectors is primarily impacted by the zenith angle \(\psi\), which is also illustrated in Fig.~\ref{fig:J_0}.

\subsection{Properties of the Unconstrained Optimal Weights}
When the RIS weights are constrained to have unit modulus, it becomes increasingly difficult to find a perfect solution that achieves deep nulls in multiple directions as the number of interference sources grows. This limitation can be understood through the energy required to cancel interference, where a perfect unimodular solution may not exist if the norm of the unconstrained optimal solution \(\mathbf{w}^*\) becomes too large. In this subsection, we investigate how the structure of \(\mathbf{A}\) affects the norm of \(\mathbf{w}^*\). Specifically, we show that as the correlation between the electric field intensity vectors increases, the condition number of \(\mathbf{A}\) increases, and so does the norm of \(\mathbf{w}^*\). \par
We begin by establishing an upper bound on the norm of the unconstrained optimal:
\begin{equation}
\Vert \mathbf{w}^*\Vert \leq  \big\Vert \mathbf{A}^H \big\Vert \bigg\Vert \left(\mathbf{AA}^H \right) ^{-1} \bigg\Vert \Vert \mathbf{y} \Vert.
\end{equation}
The following lemma shows how the off-diagonal elements of \(\mathbf{AA}^H\) affect the eigenvalue spread: 
\begin{lemma}
Given that all \(\mathbf{e}_i\) have the same norm regardless of direction and frequency, the eigenvalues \(\lambda_i\) of \(\mathbf{AA}^H\) satisfy: (Proof provided in Appendix~\ref{app:B})
\begin{equation}
\begin{split}
&\Vert \mathbf{e}_{i}\Vert^2 - \Big\Vert \mathbf{AA}^H - \mathrm{diag}\left(\mathbf{AA}^H \right) \Big\Vert_\mathrm{F} \leq \lambda_i \left(\mathbf{\mathbf{AA}}^H \right) \leq \\
&\qquad \Vert \mathbf{e}_{i}\Vert^2 + \Big\Vert \mathbf{AA}^H - \mathrm{diag}\left(\mathbf{AA}^H \right) \Big\Vert_\mathrm{F}, \ \forall \ i.
\end{split}
\end{equation}
\end{lemma}
This result implies that as the off-diagonal terms in \(\mathbf{AA}^H\) grow (i.e., as the correlation between the electric field intensity vectors increases), the eigenvalue spread widens, resulting in a larger condition number and spectral norm \(\big\Vert \mathbf{A}^H \big\Vert\). In Appendix~\ref{app:C}, we show an example of the correlation coefficient between the electric field intensity vectors versus the condition number for \(K =2\) case. \par
Next, consider the eigendecomposition of \(\mathbf{AA}^H\):
\begin{equation}
\bigg\Vert \left( \mathbf{AA}^H \right)^{-1} \bigg\Vert = \big\Vert \mathbf{V} \boldsymbol{\Lambda}^{-1} \mathbf{V}^H \big\Vert = \big\Vert \boldsymbol{\Lambda}^{-1} \big\Vert,
\label{eq:eigen_decomp}
\end{equation}
where \(\boldsymbol{\Lambda}\) is a diagonal matrix whose entries are the eigenvalues of \(\mathbf{AA}^H\) and \(\mathbf{V}\) is a unitary matrix containing the corresponding eigenvectors. Equation~(\ref{eq:eigen_decomp}) indicates that the norm of \( \left( \mathbf{AA}^H \right)^{-1}\) grows as the smallest eigenvalue of \(\mathbf{AA}^H\) shrinks (i.e., the electric field intensity vectors become more correlated). \par
Lastly, the norm of \(\mathbf{y}\) becomes large as we increase the number of interference directions to null. For a fixed number of null directions, we observe from the empirical results that \(\Vert \mathbf{y} \Vert\) remains relatively stable regardless of the interference direction and frequency. Therefore, we conclude that the norm of optimal weights \(\mathbf{w}^*\) grows large as the correlation between different electric field intensity vectors increases, which is determined by the number of independent interference sources to null, and their angle and frequency separation. A large \(\Vert \mathbf{w}^* \Vert\) makes a uniomdular solution that forms perfect nulls in all desired directions infeasible, since there is not enough energy to cancel the received energy in the multiple closely aligned interference directions. It is worth noting that reducing the reconfigurable area of the reflector increases the average radial distance of the reconfigurable rim \(\bar{\rho}\). (\ref{eq:corr_coef_cases}) indicates that such an increase leads to stronger correlation in the electric field intensity vector, which in turn makes it more challenging to obtain a unimodular solution that yields deep nulls. Conversely, a larger reconfigurable area on the dish provides more degrees of freedom, which intuitively makes it easier to achieve an optimal unimodular solution. 

\subsection{Nulling Limits Analysis}
For unconstrained complex weights, the element requirement follows from a linear degrees-of-freedom argument: with \(K\) nulling constraints and one mainlobe constraint, a solution exists when \(\mathbf{A}\) has full row rank and the number of RIS elements satisfies \(N\ge K+1\). This full-row-rank condition is generally satisfied when the interference signals come from distinct directions and/or frequencies as indicated in Section~\ref{sec:IV-A}. For unimodular weights, however, it is generally difficult to predict whether perfect nulls can be achieved at all target angles without explicitly solving the problem via iterative algorithms due to the non-convex nature of the unimodular constraint on the weights. Thus, for this case, the nulling limit cannot be characterized solely by the number of RIS elements.  \par
However, predicting whether perfect nulls can be achieved at all target angles can still be made by examining the properties of the unconstrained optimal solution \(\mathbf{w}^*\) from an energy requirement perspective. Since \(\mathbf{w}^*\) is the minimum-norm solution, it provides a lower bound on the amount of weight energy required to satisfy the nulling constraints. Therefore, a necessary condition for perfect unimodular nulling is
\begin{equation}
\Vert \mathbf{w}^* \Vert_2 \leq \sqrt{N},
\label{eq}
\end{equation}
where \(\sqrt{N}\) is the Euclidean norm of the unimodular solution. However, this bound is loose, since it only constrains the total energy of the unconstrained solution and does not account for how that energy is distributed across individual RIS elements. To obtain a more informative feasibility indicator based on the unconstraint optimal solution, we further refine this energy-requirement perspective by comparing the aggregate magnitude of the entries of \(\mathbf{w}^*\) below unit magnitude with that of the entries above unit magnitude. Specifically, a unimodular solution can be viewed as a phase-adjusted approximation to this unconstrained optimum, where each element has unit magnitude. If the number of elements in the weight vector is sufficiently large, we can model the entries of \(\mathbf{w}^*\) as random variables with varying magnitudes and phases. In this case, a unimodular solution is expected to exists only when the ``energy" in the entries of \(\mathbf{w}^*\) with magnitudes less than one dominates that of the entries with magnitude greater than one, that is
\begin{equation}
\sum_{\mathcal{I}} |w_{\mathcal{I}}^*| \geq \sum_{\mathcal{J}} |w_{\mathcal{J}}^*|,
\label{eq:criterion}
\end{equation}
where \(\mathcal{I} = \left\{n \ \big\rvert \ |w_n^*| < 1 \right\}\) and \(\mathcal{J} = \left\{n \ \big\rvert \ |w_n^*| > 1 \right\}\). Furthermore, based on our numerical observations, after sorting the entries of \(\mathbf{w}^*\) in an ascending order of magnitude, we observe that their magnitudes often follow an approximately increasing profile. For analytical tractability, we approximate this ordered profile by a linearly increasing sequence from zero to \(\Vert \mathbf{w}^* \Vert_{\infty}\). This approximation is used to convert the empirical energy-balance condition in (\ref{eq:criterion}) into a simple feasibility indicator. Specifically, define \(\epsilon = \Vert \mathbf{w}^*\Vert_{\infty}\) so that \(\epsilon/N\) is the separation of magnitudes. By making equation (\ref{eq:criterion}) an equality, we obtain
\begin{equation}
\begin{split}
\sum_{n=1}^{\lfloor N/\epsilon \rfloor} \frac{\epsilon}{N} n \: &= \sum_{n= \lfloor N/\epsilon \rfloor + 1}^{N} \frac{\epsilon}{N} n,
\end{split}
\end{equation}
where \(\lfloor \cdot \rfloor\) denotes the floor operation. Solving and taking the non-negative solution results in
\begin{equation}
\begin{split}
\epsilon &= \frac{\sqrt{2N^2 + 2N + 1} + 1}{N + 1},  \\
% &= \frac{N \sqrt{2 + \frac{2}{N} + \frac{1}{N^2}} + 1}{N + 1}, \\
\lim_{N \rightarrow \infty}\epsilon &= \sqrt{2}.
\end{split}
\end{equation}
In the next section, we will show that even though the sorted magnitude of \(\mathbf{w}^*\) is not strictly linearly increasing, this threshold provides an accurate criterion for quickly determining the existence of a unimodular solution that gives perfect nulls. \par

\section{Numerical Evaluation}
\label{sec:V}

\begin{figure}[t!]
\begin{center}
\includegraphics[scale = 0.22]{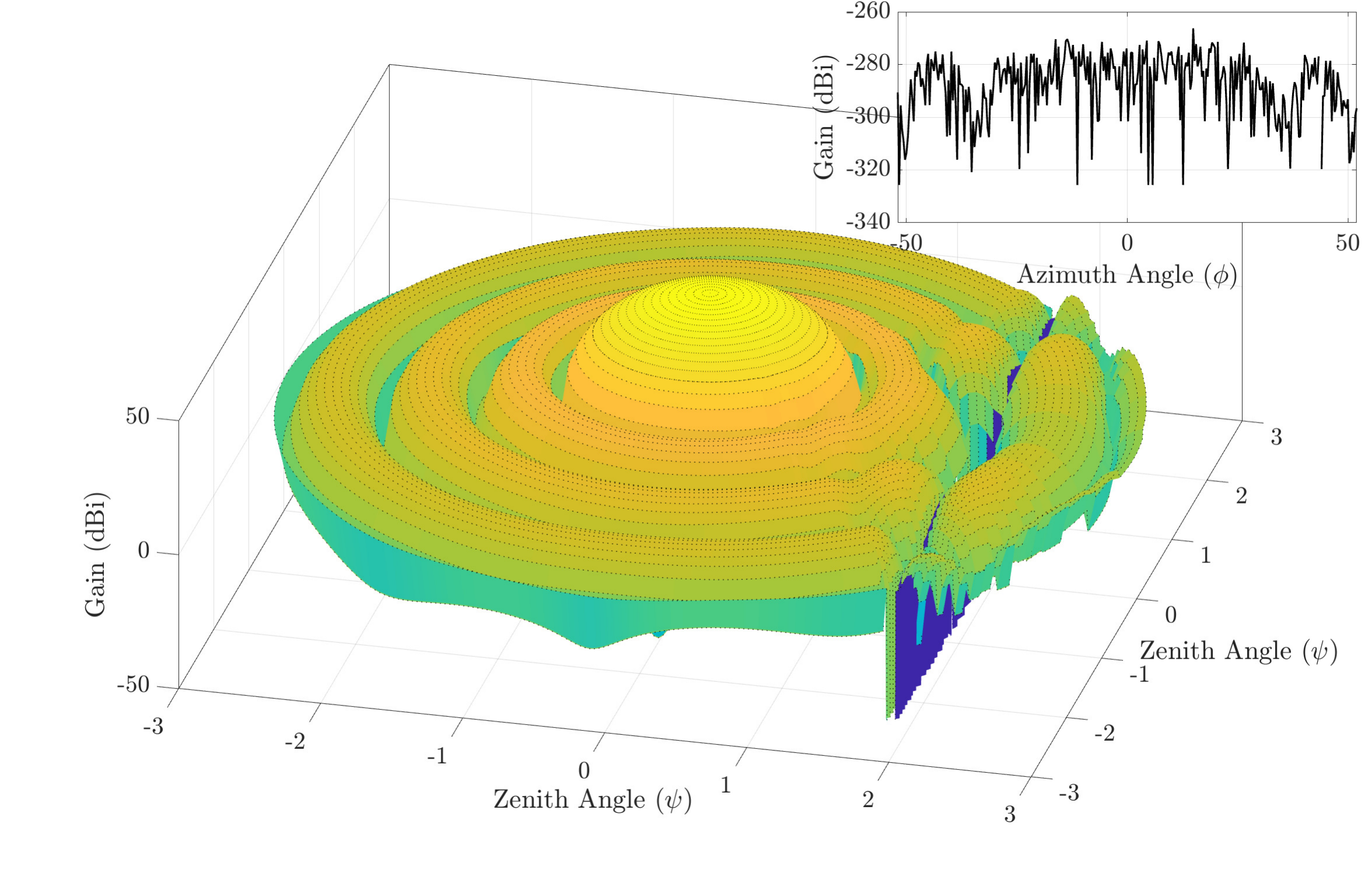}
\caption{3-D co-pol pattern for the reconfigurable 18m dish with 0.5m reconfigurable rim. The satellites passes through the second peak of the sidelobes. Continuous-phase unimodular weight vectors are applied. The subfigure on the top right shows the null depth along the satellite’s trajectory. }
\label{fig: 3D_pattern}
\end{center}
\end{figure}

\begin{figure}[t!]
\begin{center}
\includegraphics[scale = 0.55]{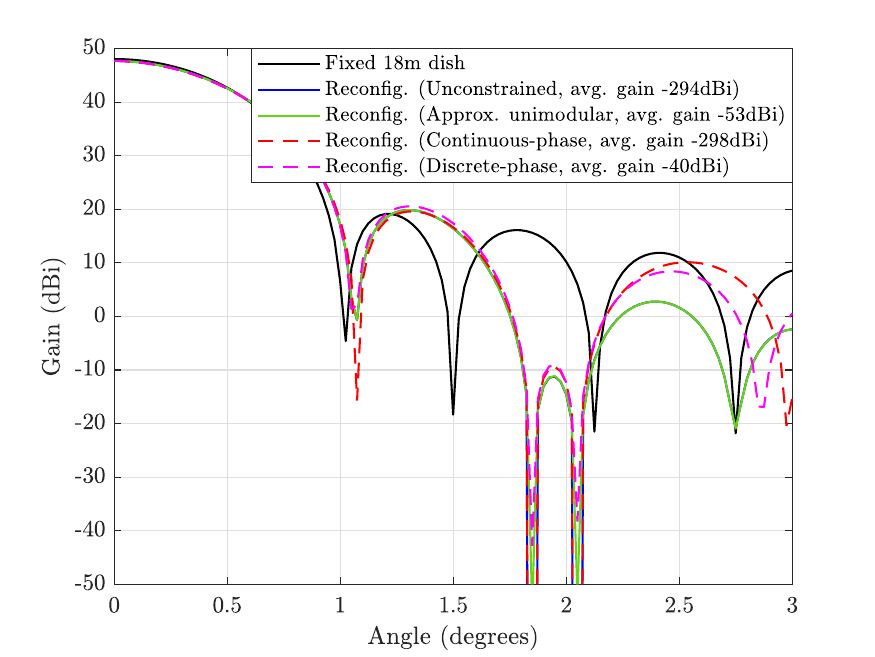}
\caption{H-plane (\(\phi = 0^\circ\)) co-pol pattern for conventional (fixed) 18m dish and reconfigurable 18m dish with nulls placed at \(\psi= 1.85^{\circ}, 2.05^{\circ}\). Quantization levels of the discrete-phase unimodular weights is set to \(M = 4\). }
\label{fig:pattern}
\end{center}
\end{figure}

To demonstrate the performance of the algorithms discussed in Section~\ref{sec:III} and to evaluate the achievable null depth for the interference direction, we present numerical results in this section. We consider an \(18\)m diameter paraboloidal reflector operating at a frequency of \(1.5\)GHz. The outer \(0.5\)m rim of the reflector is reconfigurable and is discretized into contiguous reflectarray unit cells with side length \(0.5\lambda\). For the given rim width and operating center frequency, this half-wavelength discretization results in \(N = 2756\) reconfigurable segments. The feed model is given by (\ref{eq:mag_field}) with \(q = 1.5\). Unless otherwise specified, the mainlobe constraint is set to \(\delta = 0.01\) for all simulations below. \par
We first examine the antenna pattern under a scenario where a satellite traverses the sidelobe region. Fig.~\ref{fig: 3D_pattern} shows a 3D surface plot of the co-polarized far-field pattern over the full angular domain, where the weights are restricted to continuous-phase unimodular, updated with an angular resolution of \(0.01^\circ\). We can see from Fig.~\ref{fig: 3D_pattern} that the sidelobe gain is significantly suppressed along the satellite's path, confirming the efficacy of creating the adaptive nulls using the RIS elements. Moreover, similar nulling behavior can be observed across angular cross sections that include the prescribed nulling directions, indicating that the suppression is not limited to a specific principle plane. \par
Next, we compare the beam patterns under different weight constraints. While the 3D pattern in Fig.~\ref{fig: 3D_pattern} illustrates the overall spatial nulling behavior over the full angular domain, one-dimensional pattern cuts provide a clearer visualization for comparing the null depths achieved by different methods. Therefore, we restrict our analysis to the H-plane (\(\phi = 0^\circ\)). Fig.~\ref{fig:pattern} presents the H-plane co-pol patterns when nulls are placed at \(\psi = 1.85^\circ\) and \(\psi = 2.05^\circ\). For all cases, the mainlobe gain is well-maintained. As expected, the unconstrained weights yield perfect nulls at the target angles, as the linear equality is strictly satisfied. For the continuous-phase unimodular case, the small number of angles to null and their relatively large angular separation result in weakly correlated electric field intensity vectors, allowing the GP/AP algorithms to find a solution that also achieves deep nulls. Moreover, the closed-form unimodular weights also provide satisfactory results. In contrast, the null depth of the discrete-phase unimodular weights is compromised due to the more restrictive phase quantization constraint. \par

\begin{figure}[t!]
\begin{center}
\includegraphics[scale = 0.45]{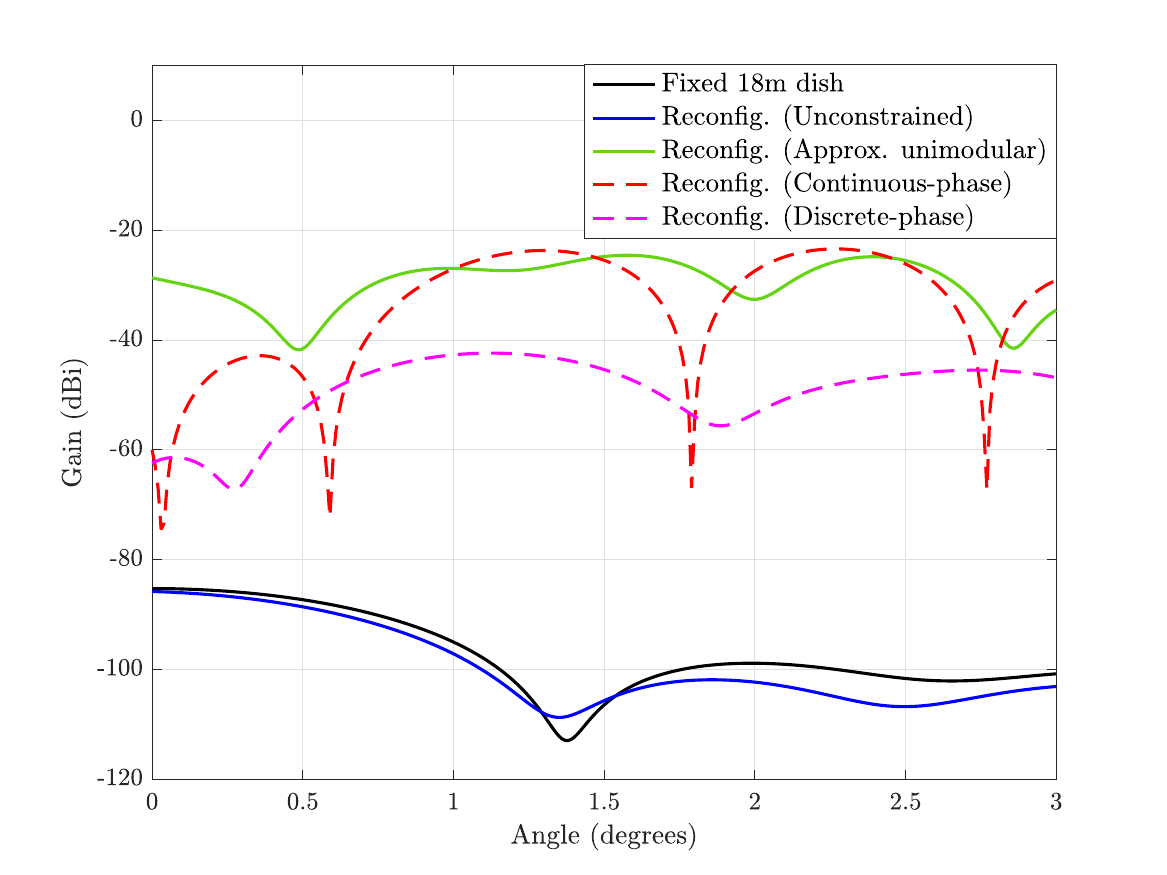}
\caption{H-plane (\(\phi = 0^\circ\)) cross-pol pattern for conventional (fixed) 18m dish and reconfigurable 18m dish with nulls placed at \(\psi= 1.85^{\circ}, 2.05^{\circ}\). Quantization levels of the discrete-phase unimodular weights is set to \(M = 4\). }
\label{fig:pattern_xpol}
\end{center}
\end{figure}

In addition to the co-pol pattern, we also examine the corresponding cross-polarized (cross-pol) pattern as an auxiliary performance metric. Although the proposed weight-design methods primarily target sidelobe nulling in the co-pol component while preserving the mainlobe response, excessive cross-pol radiation would be undesirable because it may degrade polarization purity and increase unintended interference in the orthogonal polarization channel. Therefore, the cross-pol pattern is evaluated to verify that the imposed nulling weights do not introduce significant polarization distortion. As shown in Fig.~\ref{fig:pattern_xpol}, for all considered weight constraints, the cross-pol level remains low over the angular region of interest, indicating that the proposed nulling designs maintain satisfactory polarization performance while achieving the desired co-pol nulls. \par
To further quantify the impact of the RIS weights on the overall antenna pattern, we also report the mainlobe peak loss and the maximum sidelobe level (MSL) in addition to the achievable null depths at the prescribed interference directions. The mainlobe peak loss is defined as the relative reduction in the mainlobe peak gain after applying the nulling weights, with respect to the 18m fixed-dish. Specifically, it is computed in linear scale as
\begin{equation}
p_{\mathrm{ML}} = 1 - \frac{G_0(\mathbf{w}^*)}{G_0'},
\end{equation}
where \(G_0(\mathbf{w}^*)\) denotes the mainlobe peak gain of the RIS-aided reflector after applying the optimized weight vector \(\mathbf{w}^*\), and \(G_0'\) denotes the mainlobe peak gain of the fixed 18m reflector. The maximum sidelobe level is used to verify that creating targeted nulls does not introduce excessive sidelobe growth outside the mainlobe region, and is expressed as 
\begin{equation}
G_{\mathrm{MSL}} = \max_{(\psi, \phi) \in \Omega_{\mathrm{SL}}} G_{(\psi, \phi)}(\mathbf{w}^*),
\end{equation}
where \(G_{(\psi, \phi)}(\mathbf{w}^*)\) denotes the gain of the RIS-aided reflector after applying the weight vector \(\mathbf{w}^*\) at angle \((\psi, \phi)\), and \(\Omega_{\mathrm{SL}}\) denotes the sidelobe angular range. These two quantities serve as auxiliary pattern-preservation metrics, whereas the gain at the prescribed null directions remains the primary performance metric for the targeted interference suppression problem considered in this work. In Fig.~\ref{fig:pattern}, the mainlobe peak loss is \(p_{\mathrm{ML}}=7.34\%\) for all four cases, which is consistent with the imposed mainlobe constraint of \(\delta=0.01\). In addition, the MSL is \(19\)dBi for the fixed \(18\)m dish, \(19.84\)dBi for the unconstrained weights and the approximate continuous-phase unimodular weights, \(19.59\)dBi for the continuous-phase unimodular weights, and \(20.53\)dBi for the discrete-phase unimodular weights. The MSL results show that the nulls can be achieved at the desired angles without causing significant sidelobe growth compared with the original fixed reflector. Since the mainlobe peak loss and MSL are intended to provide auxiliary pattern-preservation information, we report these quantities for the representative beam-pattern comparison in Fig.~\ref{fig:pattern}. \par
\begin{figure}[t!]
\begin{center}
\includegraphics[scale = 0.45]{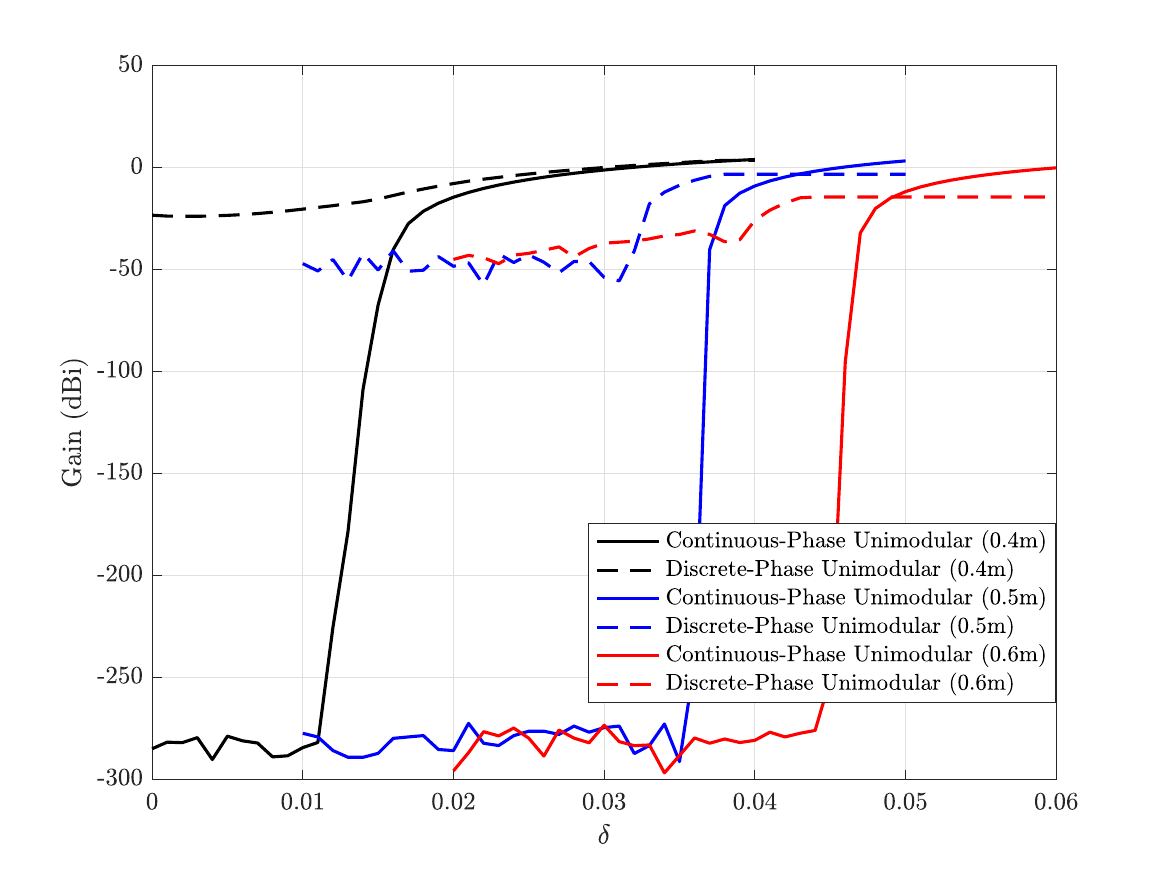}
\caption{Trade-off between the mainlobe constraint \(\delta\) and the achievable null place at \(\psi = 1.85^\circ\). Rim widths of \(0.4, 0.5, 0.6\)m correspond to \(N = 2222, 2756, 3296\) RIS elements, respectively. The quantization level is \(M = 4\). }
\label{fig:delta_vs_gain}
\end{center}
\end{figure}
While the results in Fig.~\ref{fig:pattern} correspond to a fixed mainlobe-preservation tolerance of \(\delta=0.01\), we further vary \(\delta\) to examine how relaxing or tightening the mainlobe constraint affects the achievable null depth when the weights are restricted to unit-modulus. This comparison is also performed for different numbers of RIS elements by varying the width of the reconfigurable rim while keeping the individual element size fixed. Specifically, rim widths of \(0.4\), \(0.5\), and \(0.6\)m correspond to \(N=2222\), \(2756\), and \(3296\) RIS elements, respectively. As shown in Fig.~\ref{fig:delta_vs_gain}, for a fixed number of RIS elements, increasing \(\delta\) imposes a more restrictive mainlobe constraint and reduces the flexibility for sidelobe cancellation, thereby resulting in shallower nulls. On the other hand, for a fixed \(\delta\), increasing the number of RIS elements provides additional degrees of freedom, which improves the achievable nulling performance while satisfying the mainlobe constraint. In this particular single-null example, when \(N=2756\), perfect nulling at the desired angle is no longer achievable once the mainlobe-preservation tolerance is tightened beyond approximately \(\delta \approx 0.035\).

\begin{figure}[t!]

\begin{subfigure}[b]{1\textwidth}
  \includegraphics[scale = 0.46]{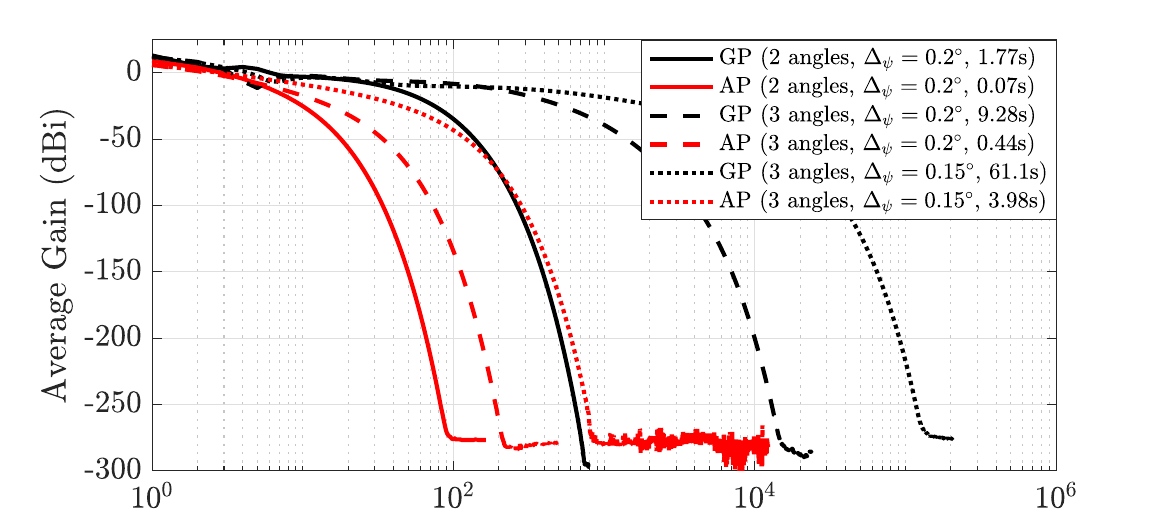}
\end{subfigure}

\medskip % insert a bit of vertical whitespace
\begin{subfigure}[b]{1\textwidth}
  \includegraphics[scale = 0.46]{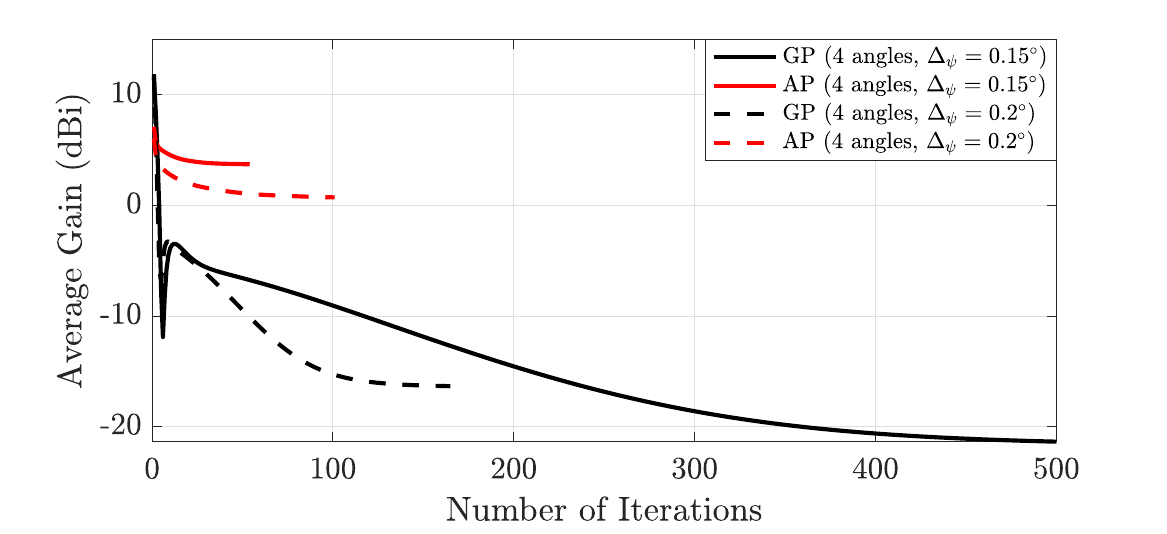}
\end{subfigure}

\caption{Comparison of convergence rate between AP and GP for finding the continuous-phase unimodular solution. The four angles to be nulled are evenly spaced with \(\Delta_\psi\) degrees between consecutive angles. Top: When the linearly equality (\ref{eq:p_multi}) can be strictly satisfied. The \(x\)-axis is shown on a log scale, and the legend includes the runtime required for convergence. Bottom: When the linearly equality (\ref{eq:p_multi}) cannot be strictly satisfied. }
\label{fig:AP_vs_GP}
\end{figure}

\begin{figure}[t!]
\begin{center}
\includegraphics[scale = 0.45]{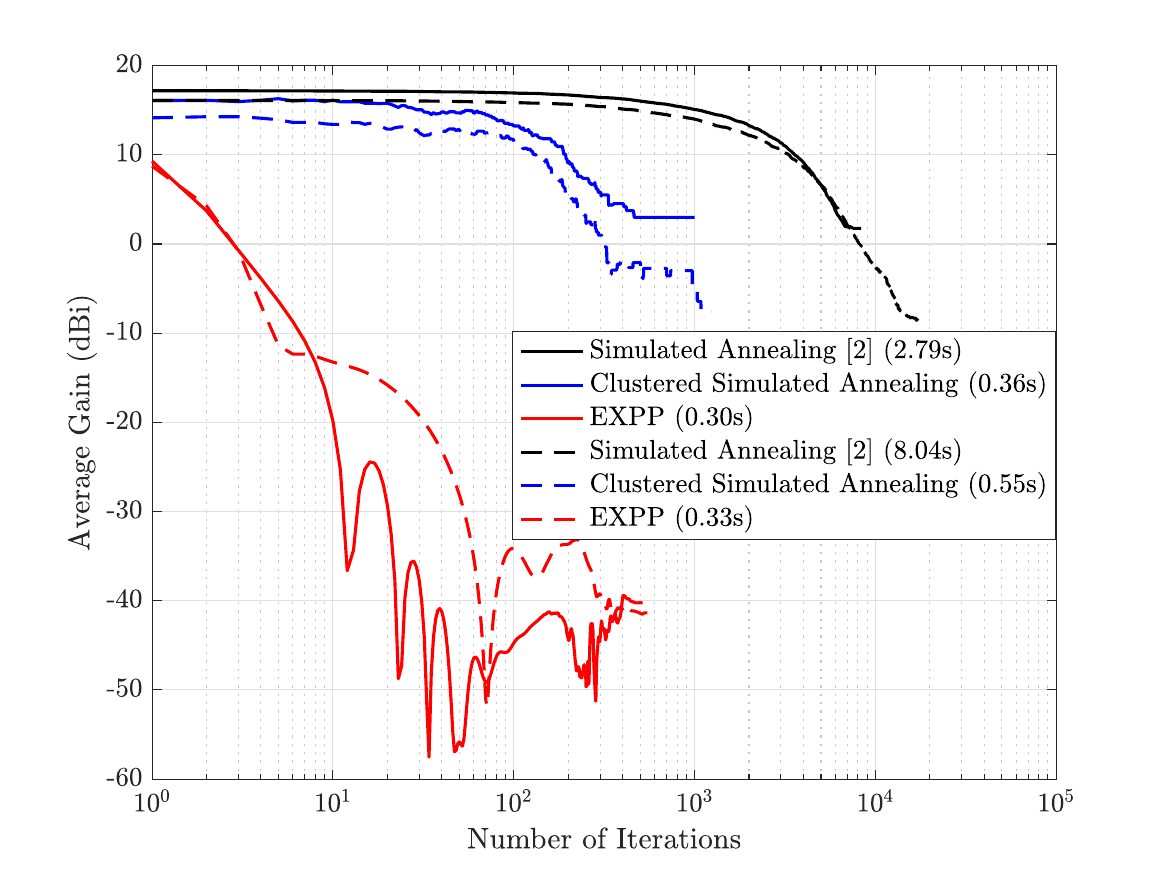}
\caption{Convergence rate of three different algorithms for finding the discrete phase unimodular solution. We group 2756 elements into 100 clusters for the clustered simulated annealing algorithm. The angles to be nulled are  \(\psi = 1.85^{\circ}\), \(2.05^{\circ}\) (solid lines) and \(\psi = 1.85^{\circ}\), \(2.05^{\circ}\), \(2.25^{\circ}\) (dashed lines). The quantization level is \(M = 4\). The \(x\)-axis is shown on a log scale, and the legend includes the runtime required for convergence. }
\label{fig:EXPP_1}
\end{center}
\end{figure}

While the previous discussion focused on the beam pattern and achievable null depths under different weight constraints, we now shift our focus to the performance comparison of the algorithms discussed in Section~\ref{sec:III}. For this analysis, we continue to restrict the far-field direction to the H-plane \(\phi = 0^{\circ}\), and only vary the zenith angle \(\psi\) to examine the impacts of the number of angles to null and the angular separation. Before comparing the convergence behavior of the iterative algorithms, we specify the stopping criteria used in the simulations. For the GP and AP algorithms, convergence is determined using an adaptive moving-window criterion based on the antenna gain in log-domain. Specifically, let \(G_{\mathrm{dB}}^{(k)} = 10\log_{10}G\left(\mathbf{w}^{(k)} \right)\) denote the gain in dB at iteration \(k\). We define the moving-window vector as
\begin{equation}
\mathbf{G}_{\mathrm{dB}}^{(k)} = \left[G_{\mathrm{dB}}^{(k-W_k+1)}, \ldots, G_{\mathrm{dB}}^{(k)} \right]^T,
\end{equation}
where \(W_k = \max\{W_{\min},\lceil \gamma k\rceil\}\) is the adaptive window length, and \(\gamma\) controls the growth rate of the window based on the iteration count. The slope \(s_k\) is obtained by LS linear regression of \(\mathbf{G}_{\mathrm{dB}}^{(k)}\) with respect to the iteration indices \([1,\ldots,W_k]^T\). Convergence is declared when the fitted gain variation over the window satisfies
\begin{equation}
|s_k|(W_k-1)\leq\epsilon
\end{equation}
for \(P_k\) consecutive iterations, where the persistence requirement \(P_k\) increases with the window length to reduce premature termination. In the simulations, the tolerance is set to \(\epsilon=0.1\). The same moving-window criterion is applied to simulated annealing and EXPP, with additional stopping conditions. Specifically, simulated annealing requires the annealing temperature to fall below a prescribed threshold, while EXPP requires the penalty parameter to reach \(c_{\max}\). The runtime of each algorithm is recorded upon satisfying its respective stopping conditions. \par
We begin by examining the performance of the GP and AP algorithms under a continuous-phase unimodular constraint when two sets \(\mathcal{S}_1\) and \(\mathcal{S}_2\) intersect, i.e., a solution exists that strictly satisfies the linear equality in (\ref{eq:p_multi}). Fig.~\ref{fig:AP_vs_GP} compares the convergence speed of GP and AP when nulling more than one interference direction, where the \(y\)-axis represents the average gain at the targeted null angles. Specifically, the gain values are first averaged over the targeted null angles in the linear domain and then converted to log domain. We can see from the top subfigure in Fig.~\ref{fig:AP_vs_GP} that AP converges faster than GP in terms of both the number of iterations and the runtime required until convergence. This advantage continues to grow as the number of interference sources increases or their angular separation decreases (increasing correlation between the electric field intensity vectors). However, \(\mathcal{S}_1\) and \(\mathcal{S}_2\) no longer intersect as the number of angles to null increases, and the resulting  performance of the AP algorithm is compromised. The bottom subfigure in Fig.~\ref{fig:AP_vs_GP} shows that the achievable gain is not as low as that of the GP algorithm, although the convergence speed is faster. In such cases, GP algorithm is preferred since it achieves better nulling performance, albeit requiring a few additional iterations to converge. \par
Next, we evaluate the EXPP algorithm against the simulated annealing algorithm given in \cite{10115669} as a representation of heuristic methods under the discrete-phase unimodular constraint. Fig. \ref{fig:EXPP_1} illustrates the convergence behavior of both algorithms and we can see that the EXPP algorithm consistently converges to a good stationary point within \(10^3\) iterations. In contrast, the simulated annealing algorithm requires over \(10^4\) iterations to converge and often fails to achieve a satisfactory solution. In order to achieve the same convergence rate using the heuristic simulated annealing algorithm, we also implemented a grouped-element strategy that updates the weights in clusters. While this grouping approach improves the convergence rate, the null depth is compromised, and the performance remains inferior to the EXPP algorithm. Therefore, based on these observations, we adopt EXPP for the subsequent simulation involving discrete-phase unimodular weights. \par

\begin{table*}[t]
\caption{Average gain for different angles on the H-plane \(\phi = 0^\circ\). }
\begin{center}
\begin{tabular}{|c|c|c|c|c|c|c|c|}
\hline
\textbf{Angles}&\multicolumn{5}{|c|}{\textbf{Average Gain (dBi)}} &  & \\
\cline{2-6} 
\textbf{(degrees)} & \textbf{Unconstrained \(\mathbf{w}^*\)} & \makecell{\textbf{Unimodular} \\ \textbf{Continuous} \\ \textbf{Phase  \(\mathbf{w}_c^*\) }} & \makecell{\textbf{Approx. Unimodular} \\ \textbf{Continuous  \(\tilde{\mathbf{w}}_c^*\) } }& \multicolumn{2}{|c|}{ \makecell{\textbf{Unimodular Discrete} \\ \textbf{Phase \(\mathbf{w}_d^*\) (\(M = 4\)) } } } &  \textbf{\textit{\(\mathrm{cond}(\mathbf{A})\)}} & \textbf{\textit{\(\mathbf{||\mathbf{w}^*||_{\infty}}\)}}\\
\cline{5-6}
& & & & \textbf{EXPP} & \textbf{Simulated Annealing} & & \\
\hline
1.85                    & -315 & -307 & -48.77 & -47.13 & -48.86 & 1.0758  & 0.7747 \\
1.85, 2.05              & -297 & -295 & -53.39 & -40.56 & 2.96   & 3.0369  & 0.8479 \\
1.85, 2.05, 2.25        & -298 & -286 & -22.21 & -42.89 & -5.80  & 11.4594 & 1.1378 \\
1.85, 2.125, 2.4, 2.675 & -300 & -287 & -5.93  & -28.15 & 1.10   & 13.5242 & 1.3923 \\
1.85, 2.1, 2.35, 2.6    & -292 & -28  & -1.80  & -22.92 & -4.29  & 18.9857 & 1.5682 \\
1.85, 2.05, 2.25, 2.45  & -296 & -23  & 2.27   & -22.17 & -14.11 & 42.0524 & 1.9420 \\
\hline
\end{tabular}
\label{tab1}
\end{center}
\end{table*}
We now demonstrate how the number and separation of interference directions affect the achievable null depth under various weight constraints. Table \ref{tab1} summarizes the achievable gains at different sets of angles (rows) with different constraints (columns). As can be seen, the unconstrained optimal weights consistently yield perfect nulls for all desired angles. However, for continuous-phase unimodular weights, nulling performance degrades when the number of nulls increases or their angular separation decreases (i.e., a higher condition number of matrix \(\mathbf{A}\) as indicated by the second last column). Moreover, the last column also confirms the validity of our conclusion in Section~\ref{sec:IV} that a perfect unimodular solution does not exist when the infinite norm of the unconstrained optimal weights is larger than approximately \(\sqrt{2}\). The third column shows the performance of the approximate unimodular solution proposed in Section~\ref{sec:apprx}. When \(\Vert \mathbf{w}^*\Vert_{\infty} \leq 1\), this approximation consistently achieves null depths near –50 dBi, which represents a substantial level of interference suppression. Given its low computational complexity, this method is particularly attractive when perfect nulling is not strictly required. For the discrete-phase unimodular case, the fifth and sixth columns compare the performance of EXPP and the simulated annealing algorithm. The performance of heuristic algorithm is significantly reduced as we increase the number of angles to null while the EXPP algorithm still yields consistent performance. \par
\begin{figure}[t!]
\begin{center}
\includegraphics[scale = 0.45]{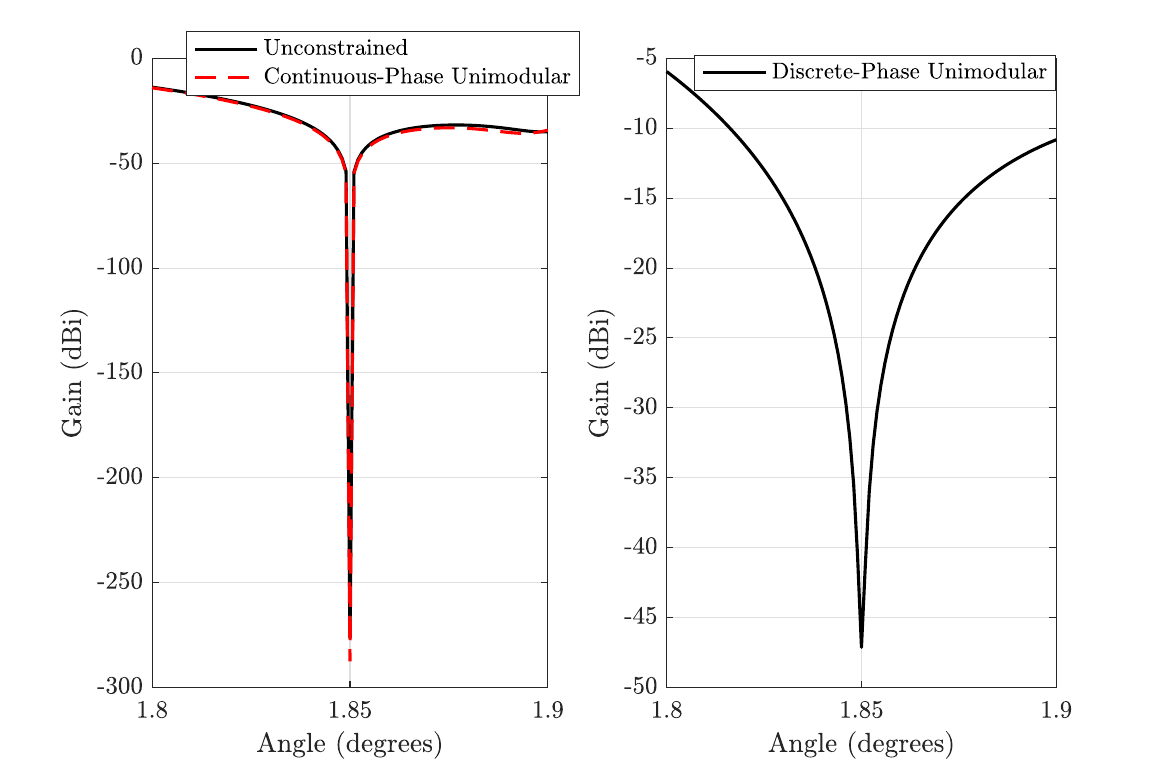}
\caption{Sensitivity of the optimized weights to angular mismatch. The weights are optimized for \(\psi = 1.85^{\circ}\). Quantization level \(M = 4\). }
\label{fig:sensiticity}
\end{center}
\end{figure}
\begin{figure}[t!]
\begin{center}
\includegraphics[scale = 0.45]{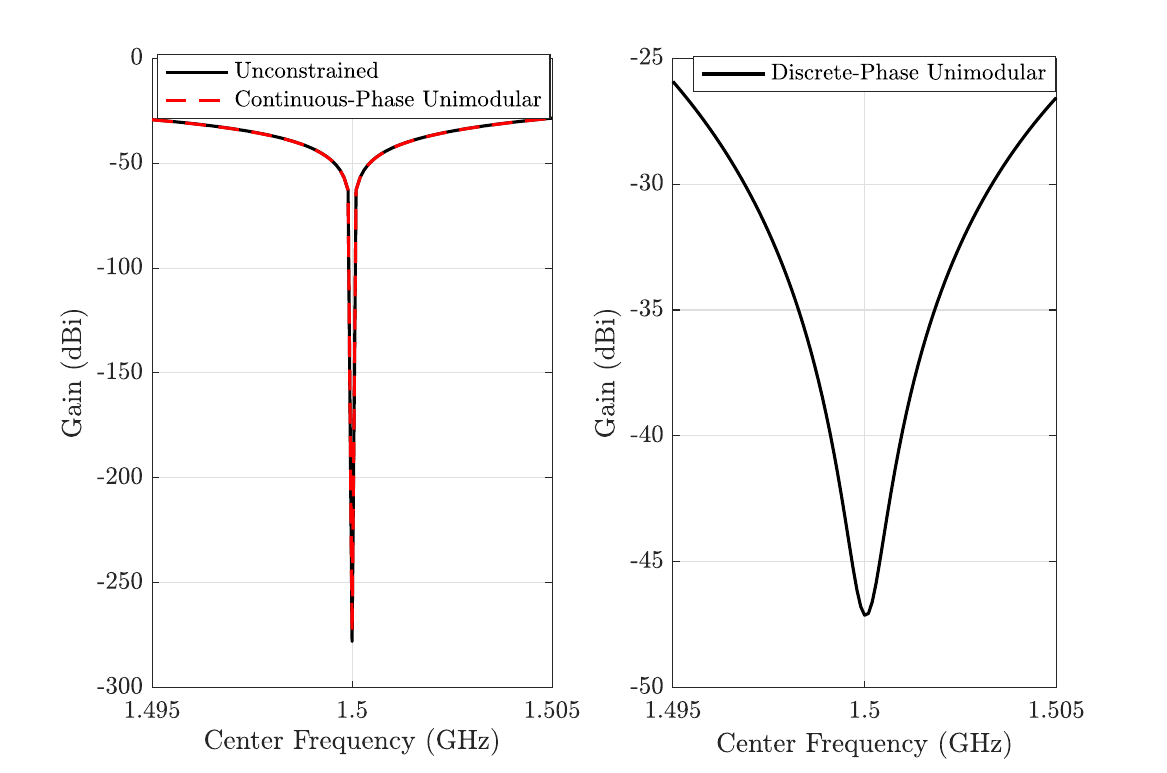}
\caption{Sensitivity of the optimized weights to center frequency mismatch. The weights are optimized for \(\psi = 1.85^{\circ}\). Quantization level \(M = 4\). }
\label{fig:freq_sensiticity}
\end{center}
\end{figure}

After evaluating the achievable null depth under perfectly specified interference directions, we further examine the robustness of the optimized weights to angle and frequency mismatch, which occurs when the angle of arrival or the center frequency of the interference signal is not estimated perfectly. Fig.~\ref{fig:sensiticity} shows the angle sensitivity for the weights under different constraints. Since the unconstrained and continuous-phase unimodular weights achieve very deep nulls at the prescribed angle, their nulling performance is highly sensitive to small angular deviations. For example, the null depth degrades from nearly \(-300\)dBi to approximately \(-50\)dBi with only a \(0.001^{\circ}\) angular shift. As the angular mismatch further increases, the degradation becomes less pronounced. In contrast, the discrete-phase unimodular weights exhibit lower sensitivity to angular mismatch, mainly because their achievable null depth at the prescribed angle is already limited by phase quantization. On the other hand, Fig.~\ref{fig:freq_sensiticity} shows the corresponding sensitivity to frequency mismatch. Similar behavior is observed: weights that produce deeper nulls at the desired frequency are more sensitive to frequency offset, whereas weights with shallower nulls exhibit less pronounced degradation. 

\section{Conclusion}
\label{sec:VI}
In this paper, we have described multiple techniques for determining optimal or near-optimal weights for creating nulls in the pattern of a prime focus-fed circular axisymmetric paraboloidal reflector antenna equipped with reconfigurable elements on the rim, serving as an example of RIS-aided spatial nulling. For unit-modulus weights, we formulate the problem as a least-squares optimization problem and solve it using projected gradient algorithms. Additionally, a closed-form solution for unimodular weights can also be found for cases where the unconstrained optimal weights have an infinite norm less than or equal to unity. For the case of discrete-phase, a MM-based algorithm was applied to solve the transformed problem which significantly outperforms the previously proposed metaheuristic method in both the convergence rate and nulling performance. We also analyze the correlation properties of the electric field intensity vectors to explain the nulling limits of the reconfigurable dish as the number of interference sources increases or their angular separation decreases. Finally, we explore the conditions under which a unimodular solution which strictly satisfies the linear equality exists based on a nulling energy requirement. Specifically, unimodular weights provide excellent performance provided that the infinite norm of the optimal unconstrained weights is less than $\sqrt{2}$. This criterion also allows us to determine when a unimodular weight solution which provides nearly perfect nulls can be found using straightforward algorithms. Simulation results validated the proposed algorithms and the theoretical limits, demonstrating their effectiveness for interference mitigation in challenging environments such as radio astronomy or spectrum-sharing under the growth of the satellite constellations. Although this work focuses on RIS-assisted spatial nulling for a single reflector receiver with reconfigurable elements on the rim, extending the proposed framework to broader wireless network settings, including RIS deployment architectures, practical implementation strategies, coordination among multiple receivers or RIS surfaces, and field-trial validation, represents an important direction for future study.

{\appendices
\section{Derivation of the Correlation Coefficient}
\label{app:A}
The derivation of the ratio \(c_n\) and the correlation coefficient \(C(\mathbf{e}_1, \mathbf{e}_2)\) is detailed in this appendix as follows. Since the interference is non-negligible for the first few sidelobes which correspond to small zenith angles \(\psi\), the co-pol direction is primarily aligned with \(\hat{\mathbf{y}}\). Under this assumption, equations (\ref{eq:mag_field}), (\ref{eq:e_discrete}) and (\ref{eq:co_pol_direction}) indicate that the \(n\)-th element of two electric field intensity vectors only differs by a phase shift, which can be written as 
\begin{equation}
\begin{split}
& c_n \approx \mathrm{exp}\left(-j(\beta_2 - \beta_1)\left( \frac{\rho_n^2}{4f}  + f\right) \right) \\
& \quad  \mathrm{exp}\bigg(j\bigg( \beta_2 \rho_n \cos(\phi_n') \sin(\psi_2)\cos(\phi_2) \\
& \qquad \quad  +\beta_2 \rho_n \sin(\phi_n') \sin(\psi_2) \sin(\phi_2)) + \frac{\rho_n^2}{4f}\beta_2 \cos(\psi_2) \\
&\qquad \quad - \beta_1 \rho_n \cos(\phi_n') \sin(\psi_1)\cos(\phi_1) \\
& \qquad \quad  - \beta_1 \rho_n \sin(\phi_n') \sin(\psi_1) \sin(\phi_1)) - \frac{\rho_n^2}{4f}\beta_1 \cos(\psi_1) \bigg) \bigg) \\
&= \mathrm{exp}\left( -j\Delta_{\beta} \left( \frac{\rho_n^2}{4f}  + f\right) \right) \\
& \quad \mathrm{exp}( j \rho_n [\cos(\phi_n') (\beta_2\sin(\psi_2)\cos(\phi_2) - \beta_1\sin(\psi_1)\cos(\phi_1))  \\
& \qquad \quad + \sin(\phi_n') (\beta_2\sin(\psi_2)\sin(\phi_2) - \beta_1\sin(\psi_1)\sin(\phi_1))]) \\
& \quad \mathrm{exp}\left( j \frac{\rho_n^2}{4f} (k_2 \cos(\psi_2) - k_1 \cos(\psi_1)) \right),\\
&\approx \mathrm{exp} \left( -j\Delta_\beta \left( \frac{\rho_n^2}{4f}   + f\right)\right) \mathrm{exp} \left( j \Delta_\beta \frac{\rho_n^2}{4f} \right) \\
& \quad  \ \mathrm{exp} (j \rho_n [\cos(\phi_n')(\beta_2 \psi_2 \cos(\phi_2) - \beta_1 \psi_1 \cos(\phi_1) ) \\
&\qquad \quad + \sin(\phi_n') (\beta_2 \psi_2 \sin(\phi_2) - \beta_1 \psi_1 \sin(\phi_1)) ]), \\
&= \mathrm{exp} (j \rho_n [\cos(\phi_n')(\beta_2 \psi_2 \cos(\phi_2) - \beta_1 \psi_1 \cos(\phi_1) ) \\
&\quad + \sin(\phi_n') (\beta_2 \psi_2 \sin(\phi_2) - \beta_1 \psi_1 \sin(\phi_1)) ])  \mathrm{exp}\left(- j \Delta_\beta f  \right)
\label{eq:ap_rn}
\end{split}
\end{equation}
where \(\Delta_\beta = \beta_2 - \beta_1\), \(\rho_n\) and \(\phi_n'\) are the radial distance and azimuth angle of the \(n\)-th element. In the approximation, we assume that \(\sin(\psi) \approx \psi\) and \(\cos(\psi_1) \approx \cos(\psi_2)\) since the angle of interest is within the first few sidelobes. Let \(a = \beta_2 \psi_2 \cos(\phi_2) - \beta_1 \psi_1 \cos(\phi_1)\), \(b = \beta_2 \psi_2 \sin(\phi_2) - \beta_1 \psi_1 \sin(\phi_1)\), and insert (\ref{eq:ap_rn}) into (\ref{eq:corr_coef_def}), the correlation between two electric field intensity vectors for different angles and frequencies can be written as
\begin{equation}
\begin{split}
C(\mathbf{e}_{1}, \mathbf{e}_{2}) &\approx \frac{1}{N} \sum_{n=1}^N e^{j \rho_n \left( \cos\left( \phi_n' \right)a  + \sin \left(\phi_n' \right) b \right) } e^{- j \Delta_\beta f}.
\label{eq:corr_coef_approx}
\end{split}
\end{equation}
By considering \(\phi' = \phi_n'\) as a random variable uniformly distributed from 0 to \(2 \pi\) and replacing \(\rho_n\) with the average radial distance for the RIS elements \(\bar{\rho} = \mathbb{E}[\rho_n]\), the expectation of the R.H.S. of (\ref{eq:corr_coef_approx}) can be expressed as 
\begin{equation}
\begin{split}
&C(\mathbf{e}_{1}, \mathbf{e}_{2}) \approx \mathbb{E}\left[ \frac{1}{N} \sum_{n=1}^Ne^{j \rho_n \left( \cos\left( \phi_n' \right)a  + \sin \left(\phi_n' \right) b \right) } e^{- j \Delta_\beta f} \right],\\
&= e^{-j \Delta_\beta f} \ \frac{1}{N} \sum_{n=1}^N \int_{0}^{2 \pi}  \frac{1}{2 \pi} e^{j \rho_n \left( \cos\left( \phi_n' \right)a  + \sin \left(\phi_n' \right) b \right) } \ \mathrm{d} \phi_n', \\
&= e^{-j \Delta_\beta f} \int_{0}^{2 \pi}  \frac{1}{2 \pi} e^{j \rho_n \left( \cos\left( \phi' \right)a  + \sin \left(\phi' \right) b \right) } \ \mathrm{d} \phi', \\
&= J_0 \left(\bar{\rho} \sqrt{a^2 + b^2} \right) e^{-j \Delta_\beta f},
\end{split}
\end{equation}
where \(J_0(\cdot)\) is the zeroth-order Bessel function of the first kind.

\section{Proof of Lemma 1}
\label{app:B}
The Courant–Fischer theorem provides the following bound for the eigenvalues of two Hermitian matrices \(\mathbf{X}\) and \(\mathbf{Y}\):
\begin{equation}
|\lambda_i(\mathbf{X}) - \lambda_i(\mathbf{Y})| \leq \parallel \mathbf{X} - \mathbf{Y} \parallel, \quad \forall \ i,
\end{equation}
In our case, let \(\mathbf{D} = \mathrm{diag}\left(\mathbf{AA}^H \right)\), which only contains the diagonal entries of \(\mathbf{AA}^H \), and \(\mathbf{E} =\mathbf{AA}^H - \mathbf{D}\), which only contains the off-diagonal entries of \(\mathbf{AA}^H\). Applying the inequality yields: 
\begin{equation}
\begin{split}
\left|\lambda_i\left(\mathbf{AA}^H \right) - \lambda_i(\mathbf{D}) \right| &\leq \big\Vert \mathbf{AA}^H - \mathbf{D}  \big\Vert = \Vert \mathbf{E} \Vert, \quad \forall \ i.
\label{eq:ineq_AA}
\end{split}
\end{equation}
From the system model, the norm of \(\mathbf{e}_i\) is identical for all directions and frequencies. Consequently, all diagonal entries of \(\mathbf{D}\) are equal to \(\Vert \mathbf{e}_i \Vert^2\) which is equal to the eigenvalue of \(\mathbf{D}\). Substituting into (\ref{eq:ineq_AA}), we obtain:
\begin{equation}
\begin{split}
\Vert \mathbf{e}_i \Vert - \lVert \mathbf{E} \rVert \leq \lambda_i \left(\mathbf{AA}^H \right) \leq \Vert \mathbf{e}_i \Vert + \lVert \mathbf{E} \rVert, \ \forall \ i
\label{eq:ineq_lambda}
\end{split}
\end{equation}
To further connect the bound with the correlation between the electric field intensity vectors, we can replace the spectral norm with the Frobenius norm using
\begin{equation}
\begin{split}
\parallel \mathbf{E} \parallel_2 &\leq \parallel \mathbf{E} \parallel_\mathrm{F} = \sqrt{\sum_i \sum_j |\mathbf{E}_{ij}|^2}.
\end{split}
\end{equation}
Therefore, (\ref{eq:ineq_lambda}) becomes
\begin{equation}
\Vert \mathbf{e}_i \Vert- \lVert \mathbf{E} \rVert_\mathrm{F} \leq \lambda_i \left(\mathbf{AA}^H \right) \leq \Vert \mathbf{e}_i \Vert + \lVert \mathbf{E} \rVert_\mathrm{F}, \ \forall \ i.
\end{equation}
This bound reveals that as the magnitude of the off-diagonal entries in \(\mathbf{AA}^H\) increases (i.e., as the correlation between different electric field intensity vectors grows), the eigenvalue spread increases.  

\section{Correlation Coefficient Versus Condition Number (\(K = 2\))}
\label{app:C}
In the case where \(K = 2\), matrix \(\mathbf{AA}^H\) can be expressed as
\begin{equation}
\mathbf{AA}^H = 
\begin{bmatrix}
\lVert \mathbf{e}_{i} \rVert^2 & \mathbf{e}_{1}^H\mathbf{e}_{2} \\
\mathbf{e}_{2}^H\mathbf{e}_{1} & \lVert \mathbf{e}_{i} \rVert^2
\end{bmatrix}.
\end{equation}
The eigenvalues of \(\mathbf{AA}^H\) can be determined by solving the equation:
\begin{equation}
\begin{split}
&\mathrm{det}\left( \mathbf{AA}^H  - \lambda I \right) = 0,\\
\lambda^2 - 2 \lVert \mathbf{e}_{i} &\rVert^2 \lambda + \lVert \mathbf{e}_{i} \rVert^4 - \left|\mathbf{e}_{\psi_1}^H\mathbf{e}_{\psi_2} \right|^2  = 0.
\label{eq:app_C_eigvalue}
\end{split}
\end{equation}
Solving (\ref{eq:app_C_eigvalue}) yields: 
\begin{equation}
\lambda_{1, 2} = \lVert \mathbf{e}_{i} \rVert^2 \pm \left| \mathbf{e}_{1}^H\mathbf{e}_{2} \right|.
\label{eq:lambda_K2}
\end{equation}
From the approximation used in (\ref{eq:corr_coef_def}), we have
\begin{equation}
\begin{split}
\mathbf{e}_{1}^H\mathbf{e}_{2} &= \sum_{n=1}^N e_{1, n}^* e_{2, n} = \sum_{n=1}^N |e_{1, n}|^2 c_n \approx N \sigma^2 C(\mathbf{e}_{1}, \mathbf{e}_{2}).
\label{eq:corr_corr_coef}
\end{split}
\end{equation}
Inserting (\ref{eq:corr_corr_coef}) in (\ref{eq:lambda_K2}), the eigenvalues can be rewritten as
\begin{equation}
\lambda_{1, 2} = \lVert \mathbf{e}_{\psi} \rVert^2 \pm N \sigma^2 C(\mathbf{e}_{1}, \mathbf{e}_{2}),
\end{equation}
which confirms the statement that a higher correlation between the electric field intensity vectors leads to a larger spread of the eigenvalues, resulting in a higher condition number. This relationship is illustrated in Fig.~\ref{fig:corr_coef_vs_condA}. 
\begin{figure}[t!]
\begin{center}
\includegraphics[width=\linewidth]{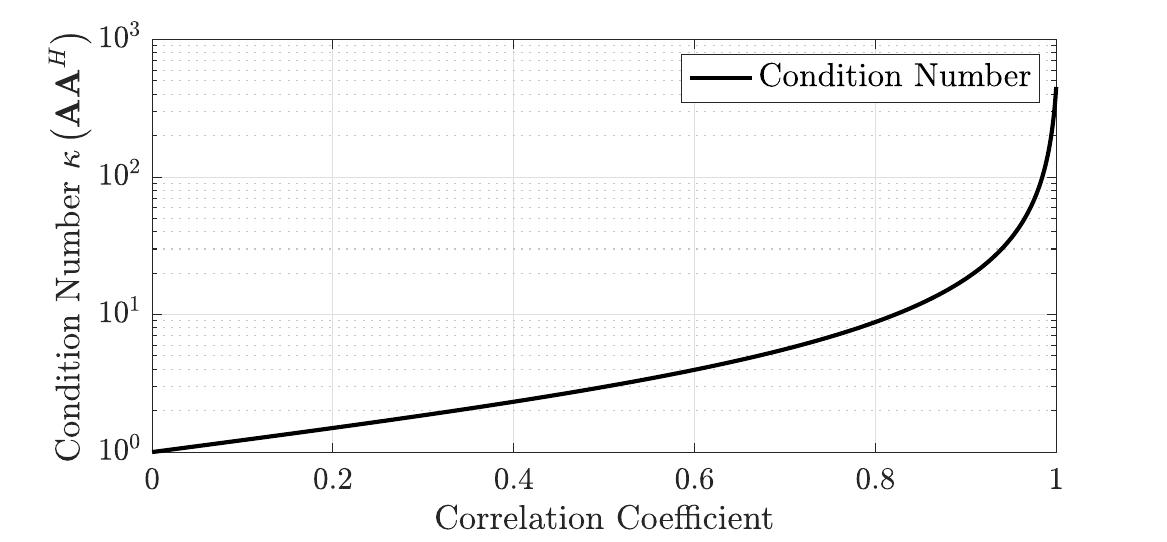}
\caption{Condition number of \(\mathbf{AA}^H\) as a function of the correlation coefficient.}
\label{fig:corr_coef_vs_condA}
\end{center}
\end{figure}

}

\bibliographystyle{IEEEtran}
\bibliography{mybibfile}

@INPROCEEDINGS{8647620,
  author={Wu, Qingqing and Zhang, Rui},
  booktitle={Proc. IEEE Global Commun. Conf. (GLOBECOM)}, 
  title={Intelligent Reflecting Surface Enhanced Wireless Network: Joint Active and Passive Beamforming Design}, 
  year={Dec. 2018},
  volume={},
  number={},
  pages={1-6},
  doi={10.1109/GLOCOM.2018.8647620}}

@INPROCEEDINGS{8485924,
  author={Tan, Xin and Sun, Zhi and Koutsonikolas, Dimitrios and Jornet, Josep M.},
  booktitle={Proc. IEEE Conf. Comput. Commun. (INFOCOM)}, 
  title={Enabling Indoor Mobile Millimeter-wave Networks Based on Smart Reflect-arrays}, 
  year={Apr. 2018},
  volume={},
  number={},
  pages={270-278},
  doi={10.1109/INFOCOM.2018.8485924}}

@INPROCEEDINGS{8683145,
  author={Wu, Qingqing and Zhang, Rui},
  booktitle={Proc. IEEE ICASSP}, 
  title={Beamforming Optimization for Intelligent Reflecting Surface with Discrete Phase Shifts}, 
  year={May 2019},
  volume={},
  number={},
  pages={7830-7833},
  doi={10.1109/ICASSP.2019.8683145}}

@INPROCEEDINGS{10001662,
  author={Zou, Zhibin and Wei, Xue and Saha, Dola and Dutta, Aveek and Hellbourg, Gregory},
  booktitle={Proc. IEEE Global Commun. Conf. (GLOBECOM)}, 
  title={{SCISRS}: Signal Cancellation using Intelligent Surfaces for Radio Astronomy Services}, 
  year={Dec. 2022},
  volume={},
  number={},
  pages={4238-4243},
  doi={10.1109/GLOBECOM48099.2022.10001662}}

@INPROCEEDINGS{8815412,
  author={Huang, Chongwen and Alexandropoulos, George C. and Yuen, Chau and Debbah, Mérouane},
  booktitle={Proc. IEEE 20th Int. Workshop Signal Process. Adv. Wireless Commun. (SPAWC)}, 
  title={Indoor Signal Focusing with Deep Learning Designed Reconfigurable Intelligent Surfaces}, 
  year={Jul. 2019},
  volume={},
  number={},
  pages={1-5},
  doi={10.1109/SPAWC.2019.8815412}}

@ARTICLE{9001052,
  author={Di, Boya and Zhang, Hongliang and Li, Lianlin and Song, Lingyang and Li, Yonghui and Han, Zhu},
  journal={IEEE Trans. Veh. Technol. }, 
  title={Practical Hybrid Beamforming With Finite-Resolution Phase Shifters for Reconfigurable Intelligent Surface Based Multi-User Communications}, 
  year={Apr. 2020},
  volume={69},
  number={4},
  pages={4565-4570},
  doi={10.1109/TVT.2020.2973202}}

@ARTICLE{9405423,
  author={ElMossallamy, Mohamed A. and Seddik, Karim G. and Chen, Wei and Wang, Li and Li, Geoffery Ye and Han, Zhu},
  journal={IEEE Trans. Veh. Technol. }, 
  title={{RIS} Optimization on the Complex Circle Manifold for Interference Mitigation in Interference Channels}, 
  year={Jun. 2021},
  volume={70},
  number={6},
  pages={6184-6189},
  doi={10.1109/TVT.2021.3073158}}

@ARTICLE{8811733,
  author={Wu, Qingqing and Zhang, Rui},
  journal={IEEE Trans. Wireless Commun. }, 
  title={Intelligent Reflecting Surface Enhanced Wireless Network via Joint Active and Passive Beamforming}, 
  year={Nov. 2019},
  volume={18},
  number={11},
  pages={5394-5409},
  doi={10.1109/TWC.2019.2936025}}

@book{monzingo2004introduction,
  title={Introduction to Adaptive Arrays},
  author={Monzingo, Robert A and Miller, Thomas W},
  year={1980},
  publisher={New York: Wiley}
}

@ARTICLE{8741198,
  author={Huang, Chongwen and Zappone, Alessio and Alexandropoulos, George C. and Debbah, Mérouane and Yuen, Chau},
  journal={IEEE Trans. Wireless Commun. }, 
  title={Reconfigurable Intelligent Surfaces for Energy Efficiency in Wireless Communication}, 
  year={Aug. 2019},
  volume={18},
  number={8},
  pages={4157-4170},
  doi={10.1109/TWC.2019.2922609}}

@ARTICLE{8013860,
  author={Chou, Hsi-Tseng and Ho, Hsien-Kwei},
  journal={IEEE Trans. Antennas Propag. }, 
  title={Local Area Radiation Sidelobe Suppression of Reflector Antennas by Embedding Periodic Metallic Elements Along the Edge Boundary}, 
  year={Oct. 2017},
  volume={65},
  number={10},
  pages={5611-5616},
  doi={10.1109/TAP.2017.2742557}}

@ARTICLE{10403767,
  author={Budhu, Jordan and Hum, Sean V. and Ellingson, Steven and Buehrer, R. Michael},
  journal={IEEE Trans. Antennas Propag.}, 
  title={Design of Rim-Located Reconfigurable Reflectarrays for Interference Mitigation in Reflector Antennas}, 
  year={Apr. 2024},
  volume={72},
  number={4},
  pages={3736-3741},
  doi={10.1109/TAP.2024.3352250}}

@INPROCEEDINGS{9886543,
  author={Buehrer, R. M. and Ellingson, S. W.},
  booktitle={Proc. IEEE Int. Symp. Antennas Propag. USNC-URSI Radio Sci. Meeting (APS/URSI)}, 
  title={Pattern Control for Reflector Antennas Using Electronically-Reconfigurable Rim Scattering}, 
  year={Jul. 2022},
  pages={577-578},
  doi={10.1109/AP-S/USNC-URSI47032.2022.9886543}}

@ARTICLE{771033,
  author={Smith, S.T.},
  journal={IEEE Trans. Signal Process.}, 
  title={Optimum phase-only adaptive nulling}, 
  year={Jul. 1999},
  volume={47},
  number={7},
  pages={1835-1843},
  doi={10.1109/78.771033}}

@ARTICLE{10284537,
  author={Vu, Trung and Raich, Raviv and Fu, Xiao},
  journal={IEEE Trans. Signal Process.}, 
  title={On Local Linear Convergence of Projected Gradient Descent for Unit-Modulus Least Squares}, 
  year={Oct. 2023},
  volume={71},
  pages={3883-3897},
  doi={10.1109/TSP.2023.3324181}}

@book{national2007handbook,
  title={Handbook of Frequency Allocations and Spectrum Protection for Scientific Uses},
  author={},
  year={2007},
  publisher={Washington, DC, USA: National Academies Press}
}

@ARTICLE{8796365,
  author={Basar, Ertugrul and Di Renzo, Marco and De Rosny, Julien and Debbah, Merouane and Alouini, Mohamed-Slim and Zhang, Rui},
  journal={IEEE Access}, 
  title={Wireless Communications Through Reconfigurable Intelligent Surfaces}, 
  year={Aug. 2019},
  volume={7},
  number={},
  pages={116753-116773},
  doi={10.1109/ACCESS.2019.2935192}}

@ARTICLE{1364142,
  author={Won-Suk Choi and Sarkar, T.K.},
  journal={IEEE Trans. Antennas Propag. }, 
  title={Phase-only adaptive Processing based on a direct data domain least squares approach using the conjugate gradient method}, 
  year={Dec. 2004},
  volume={52},
  number={12},
  pages={3265-3272},
  doi={10.1109/TAP.2004.836410}}

@ARTICLE{7547360,
  author={Sun, Ying and Babu, Prabhu and Palomar, Daniel P.},
  journal={IEEE Trans. Signal Process. }, 
  title={Majorization-{Minimization} Algorithms in Signal Processing, Communications, and Machine Learning}, 
  year={Feb. 2017},
  volume={65},
  number={3},
  pages={794-816},
  doi={10.1109/TSP.2016.2601299}}

@book{escalante2011alternating,
  title={Alternating Projection Methods},
  author={Escalante, Ren{\'e} and Raydan, Marcos},
  year={2011},
  volume={8},
  publisher={Society for Industrial and Applied Mathematics}
}

@book{russell1995modern,
  title={A modern approach},
  author={Russell, Stuart and Norvig, Peter and Intelligence, Artificial},
  journal={Artificial Intelligence. Prentice-Hall, Egnlewood Cliffs},
  volume={25},
  number={27},
  pages={79--80},
  year={1995},
  publisher={Pearson}
}

@INPROCEEDINGS{9148961,
  author={Abeywickrama, Samith and Zhang, Rui and Yuen, Chau},
  booktitle={Proc. IEEE Int. Conf. Commun. (ICC)}, 
  title={Intelligent Reflecting Surface: Practical Phase Shift Model and Beamforming Optimization}, 
  year={Jun. 2020},
  volume={},
  number={},
  pages={1-6},
  doi={10.1109/ICC40277.2020.9148961}}

@ARTICLE{10677490,
  author={Liu, Junbin and Liu, Ya and Ma, Wing-Kin and Shao, Mingjie and So, Anthony Man-Cho},
  journal={IEEE Trans. Signal Process. }, 
  title={Extreme Point Pursuit—Part {I}: A Framework for Constant Modulus Optimization}, 
  year={2024},
  volume={72},
  number={},
  pages={4541-4556},
  doi={10.1109/TSP.2024.3458008}}

@INPROCEEDINGS{10773626,
  author={Li, Xinrui and Buehrer, R. Michael and Ellingson, Steven W.},
  booktitle={MILCOM 2024 - 2024 IEEE Military Communications Conference (MILCOM)}, 
  title={An Improved Weight Selection Algorithm for Interference Mitigation in Paraboloidal Reflector Antennas with Reconfigurable Rim Scattering}, 
  year={2024},
  volume={},
  number={},
  pages={1-7},
  doi={10.1109/MILCOM61039.2024.10773626}}

@ARTICLE{9400735,
  author={Ellingson, Steven and Sengupta, Ramonika},
  journal={IEEE Antennas Wireless Propag. Lett.}, 
  title={Sidelobe Modification for Reflector Antennas by Electronically Reconfigurable Rim Scattering}, 
  year={Jun. 2021},
  volume={20},
  pages={1083-1087},
  doi={10.1109/LAWP.2021.3072536}}

@INPROCEEDINGS{10115669,
  author={Buehrer, R. Michael and Ellingson, Steven W.},
  booktitle={Proc. IEEE Aerosp. Conf.}, 
  title={Weight Selection for Pattern Control of Paraboloidal Reflector Antennas with Reconfigurable Rim Scattering}, 
  year={Mar. 2023},
  pages={1-8},
  doi={10.1109/AERO55745.2023.10115669}}

@ARTICLE{5447068,
  author={Luo, Zhiquan and Ma, Wingkin and So, Anthony Mancho and Ye, Yinyu and Zhang, Shuzhong},
  journal={IEEE Signal Process. Mag.}, 
  title={Semidefinite Relaxation of Quadratic Optimization Problems}, 
  year={May 2010},
  volume={27},
  number={3},
  pages={20-34},
  doi={10.1109/MSP.2010.936019}}

@ARTICLE{7849224,
  author={Tranter, John and Sidiropoulos, Nicholas D. and Fu, Xiao and Swami, Ananthram},
  journal={IEEE Trans. Signal Process.}, 
  title={Fast Unit-Modulus Least Squares With Applications in Beamforming}, 
  year={Jun. 2017},
  volume={65},
  number={11},
  pages={2875-2887},
  doi={10.1109/TSP.2017.2666774}}

@ARTICLE{8811616,
  author={Shao, Mingjie and Li, Qiang and Ma, Wing-Kin and So, Anthony Man-Cho},
  journal={IEEE Trans. Signal Process.}, 
  title={A Framework for One-Bit and Constant-Envelope Precoding Over Multiuser Massive {MISO} Channels}, 
  year={Oct. 2019},
  volume={67},
  number={20},
  pages={5309-5324},
  doi={10.1109/TSP.2019.2937280}}

@book{stutzman2012antenna,
  title={Antenna Theory and Design},
  author={Stutzman, Warren L and Thiele, Gary A},
  year={2012},
  publisher={NewYork, NY, USA: Wiley}
}

@ARTICLE{9681803,
  author={Jiang, Tao and Yu, Wei},
  journal={IEEE J. Sel. Areas Commun. }, 
  title={Interference Nulling Using Reconfigurable Intelligent Surface}, 
  year={May 2022},
  volume={40},
  number={5},
  pages={1392-1406},
  doi={10.1109/JSAC.2022.3143220}}

@article{alexandropoulos2023ris,
  title={{RIS}-enabled smart wireless environments: Deployment scenarios, network architecture, bandwidth and area of influence},
  author={Alexandropoulos, George C and others},
  journal={EURASIP J. Wireless Commun. Netw. },
  volume={2023},
  number={1},
  pages={103},
  year={Oct. 2023},
  publisher={Springer}
}

@ARTICLE{10172310,
  author={Ren, Yanqing and others},
  journal={IEEE Wireless Commun. Lett. }, 
  title={On Deployment Position of {RIS} in Wireless Communication Systems: Analysis and Experimental Results}, 
  year={Oct. 2023},
  volume={12},
  number={10},
  pages={1756-1760},
  doi={10.1109/LWC.2023.3292125}}

\vspace{11pt}

\begin{IEEEbiography}
[{\includegraphics[width=1in,height=1.25in,keepaspectratio]{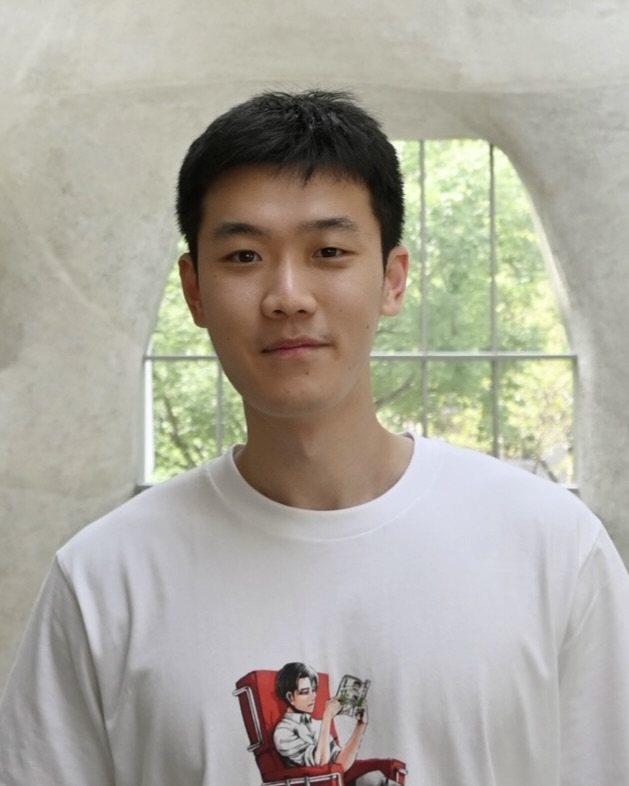}}]
{Xinrui Li} 
received the B.S. and M.S. degrees in electrical engineering from Virginia Tech, Blacksburg, VA, in 2019 and 2025, respectively, where he is pursuing the Ph.D. degree with the Bradley Department of Electrical and Computer Engineering. His research interests span the areas of communication theory and reconfigurable intelligent surface. 

\end{IEEEbiography}%

\begin{IEEEbiography}
[{\includegraphics[width=1in,height=1.25in,keepaspectratio]{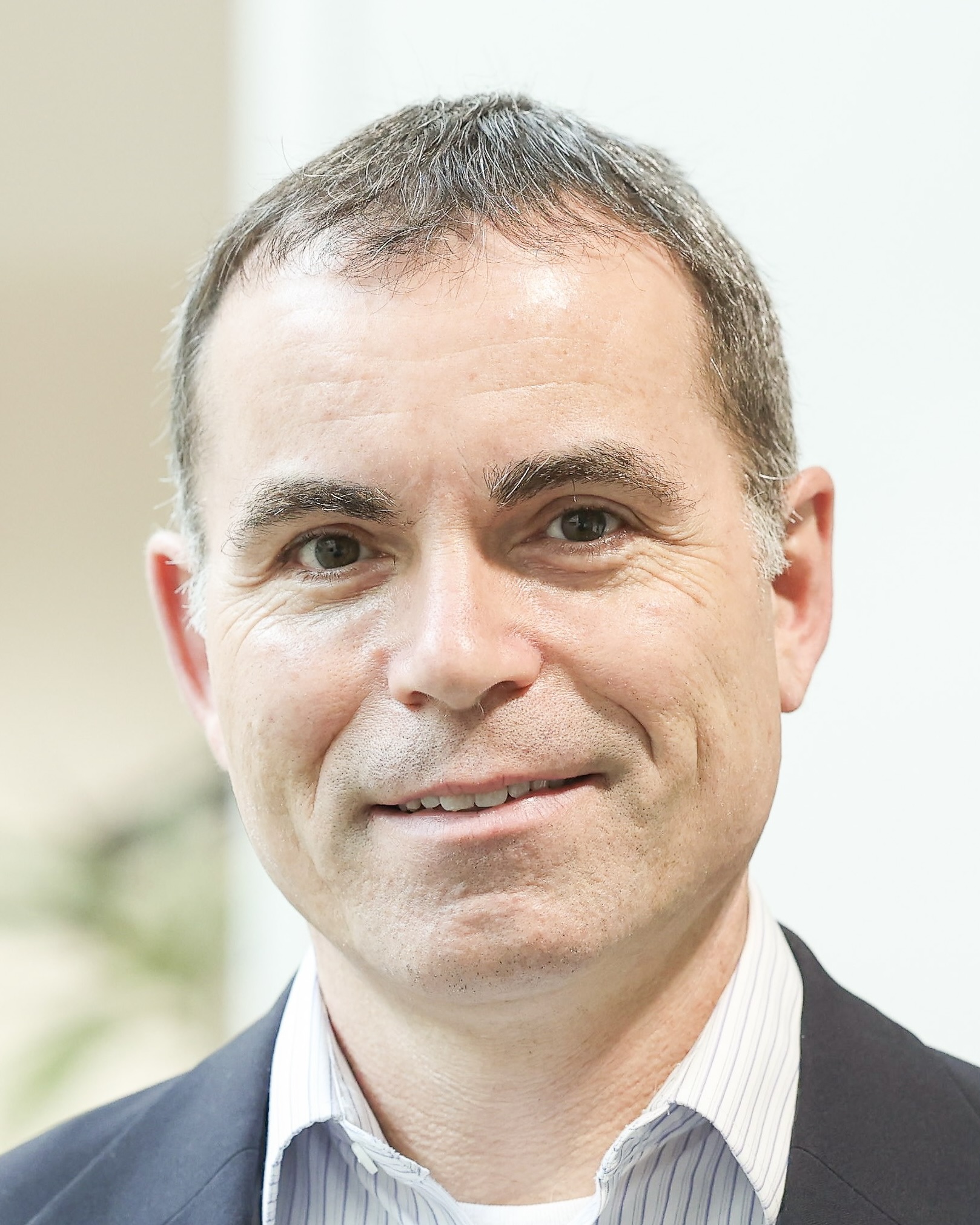}}]
{R. Michael Buehrer}{\space}(Fellow, IEEE) joined Virginia Tech from Bell Labs as an Assistant Professor with the Bradley Department of Electrical and Computer Engineering in 2001. He is currently a Professor and the Fred C. Lee Endowed Chair of Electrical and Computer Engineering.  He is also the former Director of Wireless~@~Virginia Tech, a comprehensive research group focusing on wireless communications, radar and localization. During 2009, Dr. Buehrer was a visiting researcher at the Laboratory for Telecommunication Sciences (LTS) a federal research lab which focuses on telecommunication challenges for national defense. While at LTS, his research focus was in the area of cognitive radio with a particular emphasis on statistical learning techniques.

Dr. Buehrer was named an IEEE Fellow in 2016 ``for contributions to wideband signal processing in communications and geolocation.” In 2023, he received the prestigious MILCOM Lifetime Award for Technical Achievement. This award recognizes individuals who have made important technical contributions to military communications over the course of their careers. His current research interests include machine learning for wireless communications and radar, geolocation, position location networks, cognitive radio, cognitive radar, electronic warfare, dynamic spectrum sharing, communication theory, Multiple Input Multiple Output (MIMO) communications, spread spectrum, interference avoidance, and propagation modeling. His work has been funded by the National Science Foundation, the Defense Advanced Research Projects Agency, the Office of Naval Research, the Army Research Office, the Air Force Research Lab and several industrial sponsors.

Dr. Buehrer has authored or co-authored over 100 journal and approximately 300 conference papers and holds 18 patents in the area of wireless communications. In 2023, he received the prestigious MILCOM Lifetime Award for Technical Achievement, an award that recognizes individuals who have made important technical contributions to military communications over the course of their careers. In 2023 and 2021 he was the co-recipient of the Vanu Bose Award for the best paper at IEEE MILCOM. In 2023 and 2010 he was co-recipient of the Fred W. Ellersick MILCOM Award for the best paper in the unclassified technical program. He was formerly an Area Editor IEEE Wireless Communications. He was also formerly an associate editor for IEEE Transactions on Communications, IEEE Transactions on Vehicular Technologies, IEEE Transactions on Wireless Communications, IEEE Transactions on Signal Processing, IEEE Wireless Communications Letters, and IEEE Transactions on Education. He has also served as a guest editor for special issues of The Proceedings of the IEEE, and IEEE Transactions on Special Topics in Signal Processing. In 2003 he was named Outstanding New Assistant Professor by the Virginia Tech College of Engineering and in 2014 he received the Dean’s Award for Excellence in Teaching.
\end{IEEEbiography}%

\begin{IEEEbiography}
[{\includegraphics[width=1in,height=1.25in,keepaspectratio]{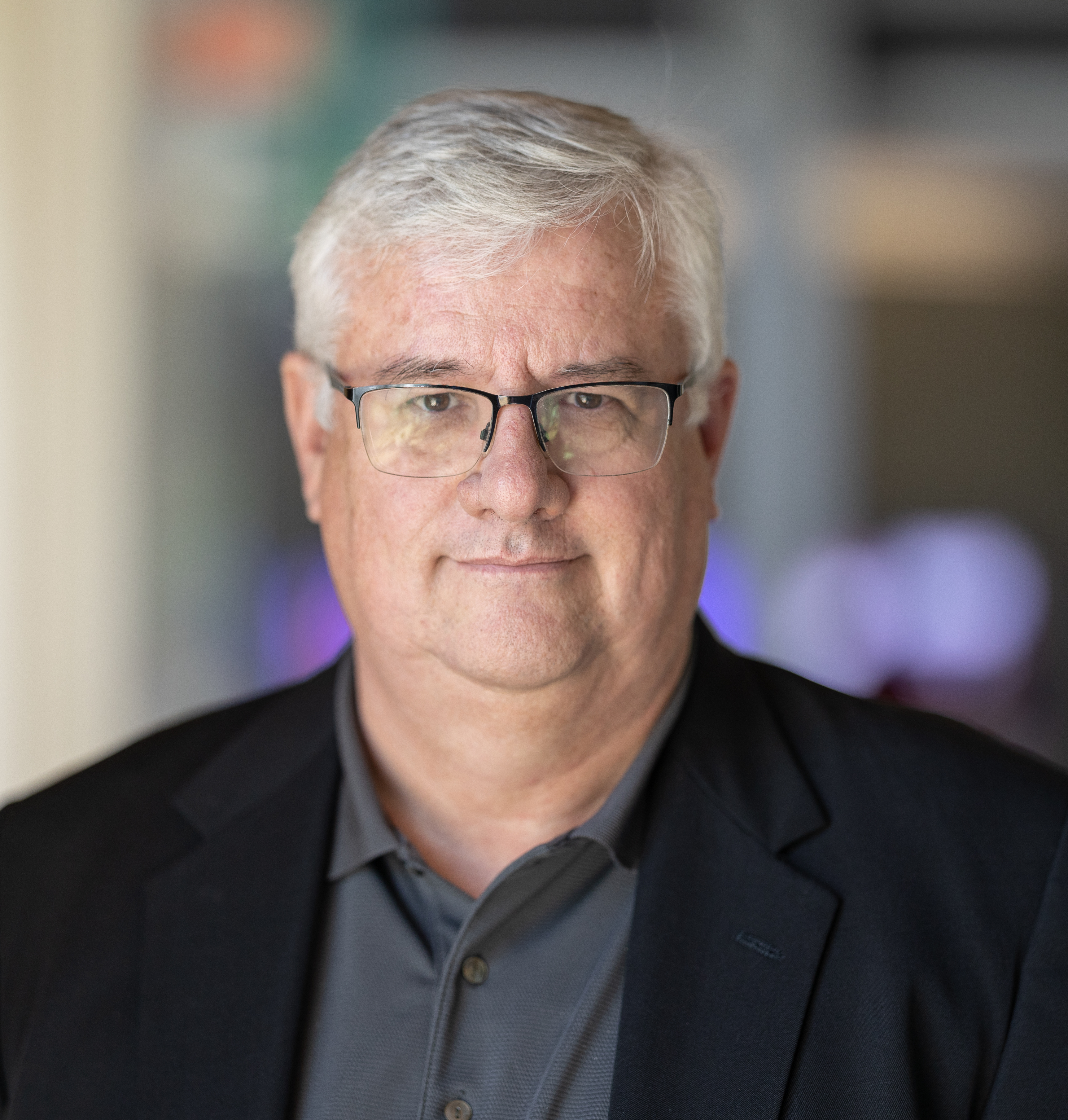}}]
{Steven W. Ellingson}{\space}(Senior Member, IEEE) (S’87-M’90-SM’03) is an Associate Professor with the Bradley Department of Electrical \& Computer Engineering at Virginia Tech, Blacksburg, VA, USA. He received the B.S. degree in electrical and computer engineering from Clarkson University, and the M.S. and Ph.D. degrees in electrical engineering from the Ohio State University. His previous positions include active duty with the U.S. Army, Senior Consultant with Booz-Allen \& Hamilton, Senior Systems Engineer with Raytheon E-Systems, and Research Scientist with the Ohio State University ElectroScience Laboratory. His research interests include antennas \& propagation, applied signal processing, and RF instrumentation.
\end{IEEEbiography}%

\vfill

\end{document}